\documentclass[%
prd,
nofootinbib,
superscriptaddress
]{revtex4}
\usepackage{amsmath,amssymb,bm,float,mathrsfs}
\usepackage{graphicx}
\usepackage{dcolumn}
\usepackage{color}
\usepackage{hyperref}
\hypersetup{
    colorlinks=true,       % false: boxed links; true: colored links
    linkcolor=red,          % color of internal links
    citecolor=blue,        % color of links to bibliography
    filecolor=magenta,      % color of file links
    urlcolor=magenta           % color of external links
}

\usepackage{comment}

\newcommand{\be}{\begin{equation}} 
\newcommand{\ee}{\end{equation}}
\newcommand{\bea}{\begin{eqnarray}} 
\newcommand{\eea}{\end{eqnarray}}

\begin{document}

%****************************** TITLE **********************************%
\title{Spectator dark matter in non-standard cosmologies}

%***************************** AUTHORS ********************************%

\author{Catarina Cosme}
\email[Email: ]{ccosme@physics.carleton.ca}
\affiliation{
Ottawa-Carleton Institute for Physics, Carleton University, 1125 Colonel By Drive, Ottawa, Ontario K1S 5B6, Canada
}
\author{Tommi Tenkanen}
\email[Email: ]{ttenkan1@jhu.edu}
\affiliation{
Department of Physics and Astronomy, Johns Hopkins University, 
Baltimore, MD 21218, USA
}

%****************************** ABSTRACT *******************************%
\begin{abstract}
It has been shown that the observed dark matter (DM) abundance can be produced by amplification of quantum fluctuations of an energetically subdominant scalar field during inflation. In this paper, we study the robustness of this ``spectator dark matter" scenario to changes in the expansion rate of the early Universe. Compared to the standard radiation-dominated (RD) scenario, two aspects will change: the DM energy density evolves differently as a function of time, and also the DM isocurvature perturbation spectrum will be different from the result in the RD case. These can impose sizeable changes to the values of model parameters which allow the field to constitute all DM while simultaneously satisfying all observational constraints. We study both free and self-interacting DM in scenarios with non-standard expansion and quantify the changes to the cases with a standard cosmological history. We also discuss testability of the scenario through primordial DM isocurvature and non-Gaussianity.
\end{abstract}

%***************************** PREPRINT ********************************%
\preprint{}

\maketitle
%%%%%%%%%%%%%%%%%%%%%%%%%%%%%%%%%%%%%%%%%%%%%%%%%%%%%%%%%%%%%%%%%%%%%%

\section{Introduction}
\label{introduction}

The deeper nature of dark matter (DM) is unknown. While the observational evidence for the existence of DM is overwhelming, its possible connection to particle physics remains poorly understood. As of today, the experimental searches of particle DM have yielded only null results \cite{Boveia:2018yeb,Hooper:2018kfv,Schumann:2019eaa}, either constraining or ruling out various models or even model paradigms of particle DM. In particular, the usual freeze-out paradigm appears less and less likely to explain the origins of DM \cite{Arcadi:2017kky}.

As the only evidence for DM are obtained through its gravitational effects -- its imprints on the Cosmic Microwave Background (CMB) and the large scale structure of the Universe, gravitational lensing and dynamics of galaxy clusters, rotational velocity curves of individual galaxies and so on --, we are motivated to ask: what if dark matter couples to the Standard Model (SM) particles only via gravity? It is clear that observationally this is not a problem, and simple and appealing mechanisms for the generation of DM have been found too. Indeed, the observed DM abundance may have been initiated purely gravitationally in the early Universe, either during or right after cosmic inflation. This idea dates back to 1980s \cite{Ford:1986sy,Turner:1987vd} and it has been gaining increasing attention recently, see e.g. Refs. \cite{Enqvist:2014zqa,Graham:2015rva,Nurmi:2015ema,Garny:2015sjg,Markkanen:2015xuw,Kainulainen:2016vzv,Bertolami:2016ywc,Heikinheimo:2016yds,Cosme:2017cxk,Enqvist:2017kzh,Cosme:2018nly,Graham:2018jyp,Alonso-Alvarez:2018tus,Ema:2018ucl,Fairbairn:2018bsw,Markkanen:2018gcw,Cosme:2018wfh,Guth:2018hsa,Ho:2019ayl,Padilla:2019fju,Tenkanen:2019aij,Tenkanen:2019wsd,AlonsoAlvarez:2019cgw,Ema:2019yrd,Laulumaa:2020pqi,Ahmed:2020fhc,Karam:2020rpa} for recent studies on gravitational DM production in different contexts.

In this paper, we will focus on a particular scenario where DM is gravitationally produced: the so-called spectator dark matter scenario, where the DM is generated by amplification of vacuum fluctuations of an energetically subdominant scalar field during inflation. The starting point is well-motivated, as weakly coupled scalar fields are typically abundant in extensions of the SM \cite{Arvanitaki:2009fg,Marsh:2015xka,Stott:2017hvl} and their dynamics during inflation is expected to provide the generic initial conditions for non-thermal production of DM after inflation \cite{Enqvist:2014zqa} -- unless the scalar field(s) themselves constitute all or part of the observed DM abundance. Previously, spectator DM scenarios have been considered in the case of free \cite{Turner:1987vd,Tenkanen:2019aij,Tenkanen:2019wsd}, self-interacting \cite{Peebles:1999fz,Nurmi:2015ema,Kainulainen:2016vzv,Heikinheimo:2016yds,Enqvist:2017kzh,Markkanen:2018gcw,Padilla:2019fju}, and non-minimally coupled cases \cite{Cosme:2017cxk,Cosme:2018nly,Cosme:2018wfh,Alonso-Alvarez:2018tus,Fairbairn:2018bsw,AlonsoAlvarez:2019cgw,Laulumaa:2020pqi}, as well as in scenarios where the DM is coupled to the inflaton \cite{Bertolami:2016ywc}, or is axion-like \cite{Beltran:2006sq,Graham:2018jyp,Guth:2018hsa,Ho:2019ayl}, but -- to the best of our knowledge -- never in the context of general non-standard (i.e. non-radiation-dominated) expansion after inflation. 

In this paper we study the robustness of the spectator dark matter scenario studied in Refs. \cite{Markkanen:2018gcw,Tenkanen:2019aij} to changes in the early Universe's expansion rate. In particular, we concentrate on scenarios where the total energy density after inflation was dominated by a perfect fluid other than radiation, for example massive meta-stable particles or a fast-rolling scalar field, or where the non-radiation-dominated expansion was caused by e.g. a period of slow reheating after inflation. All these possibilities are well-motivated; for an extensive review of such scenarios, see Ref. \cite{Allahverdi:2020bys}. Compared to the standard radiation-dominated (RD) scenario, in the context of non-standard expansion two aspects in the spectator DM model will be different: the DM energy density will evolve differently as a function of time, and also the dependence of the DM perturbation spectrum on the initial spectator field value will be different from the result in the usual RD case. 

These can impose sizeable changes to the values of model parameters which allow the field to constitute all DM while simultaneously satisfying all observational constraints. Here we study both free and self-interacting DM within a non-standard expansion and quantify the changes to the cases with a standard cosmological history. We will also discuss testability of the scenario through primordial DM isocurvature and non-Gaussianity, highlighting the fact that even though DM may couple to ordinary matter only via gravity, it does not mean that the scenario would not be testable.

The paper is organized as follows: in Sec. \ref{inflation}, we present the model and discuss the field dynamics during inflation, whereas Sec. \ref{dynamics} is devoted for the dynamics after inflation. In Sec. \ref{perturbations}, we compute the DM perturbation spectrum and investigate how the scenario can be contrasted with CMB observations. In Sec. \ref{results}, we present our results and discuss testability of the model. Finally, in Sec. \ref{conclusions}, we conclude with a brief outlook.

%%%%%%%%%%%%%%%%%%%%%%%%%%%%%%%%%%%%%%%%%%%%%%%%%%%%%%%%%%%%%%%%%%%%%%%%%%%%%%%%%%%%%%%%%%%%%%%

\section{Scalar field evolution during inflation}
\label{inflation}

We begin by reviewing the standard treatment for the evolution of a light scalar field during inflation. For dark matter, we consider the Lagrangian
\begin{equation}
\label{lagrangian}
\mathcal{L}_\chi=\frac12\partial^\mu\chi\partial_\mu\chi - V(\chi)\,,
\end{equation}
where
\begin{equation}
\label{potential}
V(\chi) =  \frac{1}{2}m^2 \chi^2 + \frac{\lambda}{4}\chi^4\,,
 \end{equation}
where $\chi$ is a real scalar field which we assume was an energetically subdominant spectator field during inflation, $m$ is its mass and $\lambda$ is a quartic self-interactions coupling. We assume that the theory is defined in a frame where the field that drives inflation, the inflaton field\footnote{The inflaton field could be, for example, the SM Higgs field \cite{Bezrukov:2007ep,Bauer:2008zj}. However, here we remain agnostic of the inflaton sector and its couplings to the SM fields and/or gravity; for reviews of Higgs-like inflation, see Refs. \cite{Rubio:2018ogq,Tenkanen:2020dge}. Likewise, we assume here that the possible couplings between the field $\chi$ and the inflaton/Higgs field are negligible and do not affect the field dynamics either during or after inflation (by e.g. rendering the $\chi$ field heavy during inflation), nor contribute to the final DM yield through any mechanism, such as production of $\chi$ quanta during reheating. For studies where these assumptions are relaxed, see e.g. Refs. \cite{Berlin:2016vnh,Adshead:2016xxj,Tenkanen:2016jic,Berlin:2016gtr}. For recent studies on scenarios where the DM field couples non-minimally to gravity, see e.g. Refs. \cite{Cosme:2017cxk,Cosme:2018nly,Cosme:2018wfh,Alonso-Alvarez:2018tus,AlonsoAlvarez:2019cgw,Laulumaa:2020pqi}.}, couples minimally to gravity and where the background during inflation space-time is well approximated by that of de Sitter, i.e. the Hubble scale during inflation is approximately constant, $H= \dot{a}/a \simeq H_{\rm inf}$. This is well motivated, as in slow-roll models of inflation that provide the best fit to data (such as plateau models, see e.g. Ref. \cite{Chowdhury:2019otk}) it is usually a very good approximation that the Hubble rate did not decrease much during inflation. Therefore, we will maintain this assumption throughout the paper. Furthermore, we assume that the inflaton field sources a major part of the primordial curvature perturbations, which eventually lead to the observed temperature fluctuations in the CMB.

\begin{figure}[H]
\begin{centering}
\includegraphics[width=0.75\textwidth]{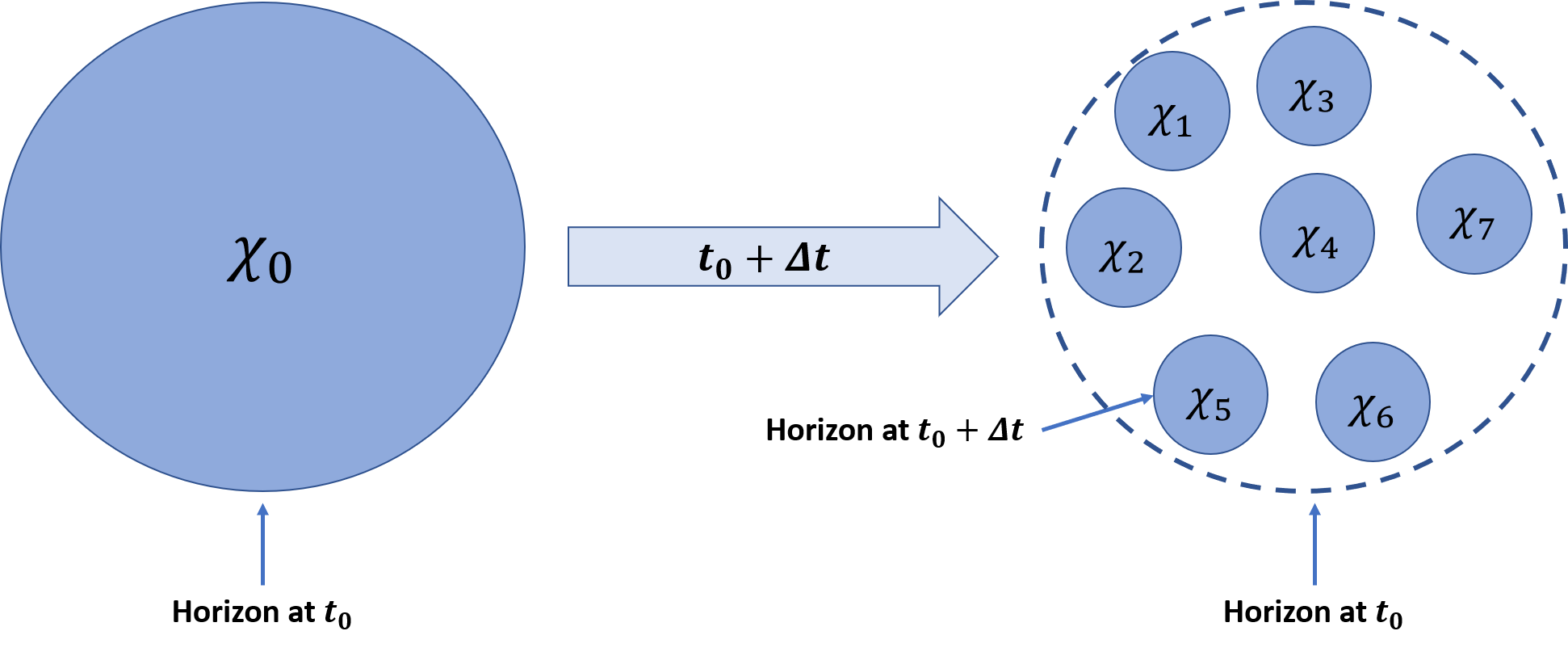}
\par\end{centering}
\caption{When a scalar field $\chi$ is light during inflation, it acquires fluctuations which get it displaced from its initial value $\chi_0$. As inflation proceeds, the (coarse-grained) scalar field performs random walk, and the Universe ends up having an ensemble of Hubble volumes, in each of which the field has a value ($\chi_{1}$,$\chi_{2}$,...) that generically differs from the average value due to the random fluctuations. The final distribution of values $P(\chi)$ is given by Eq. \eqref{eq:p}. See also Fig. \ref{distributions}.
}
 \label{regions}
\end{figure}

\begin{figure}[H]
\label{distributions}
\begin{centering}
\includegraphics[width=0.475\textwidth]{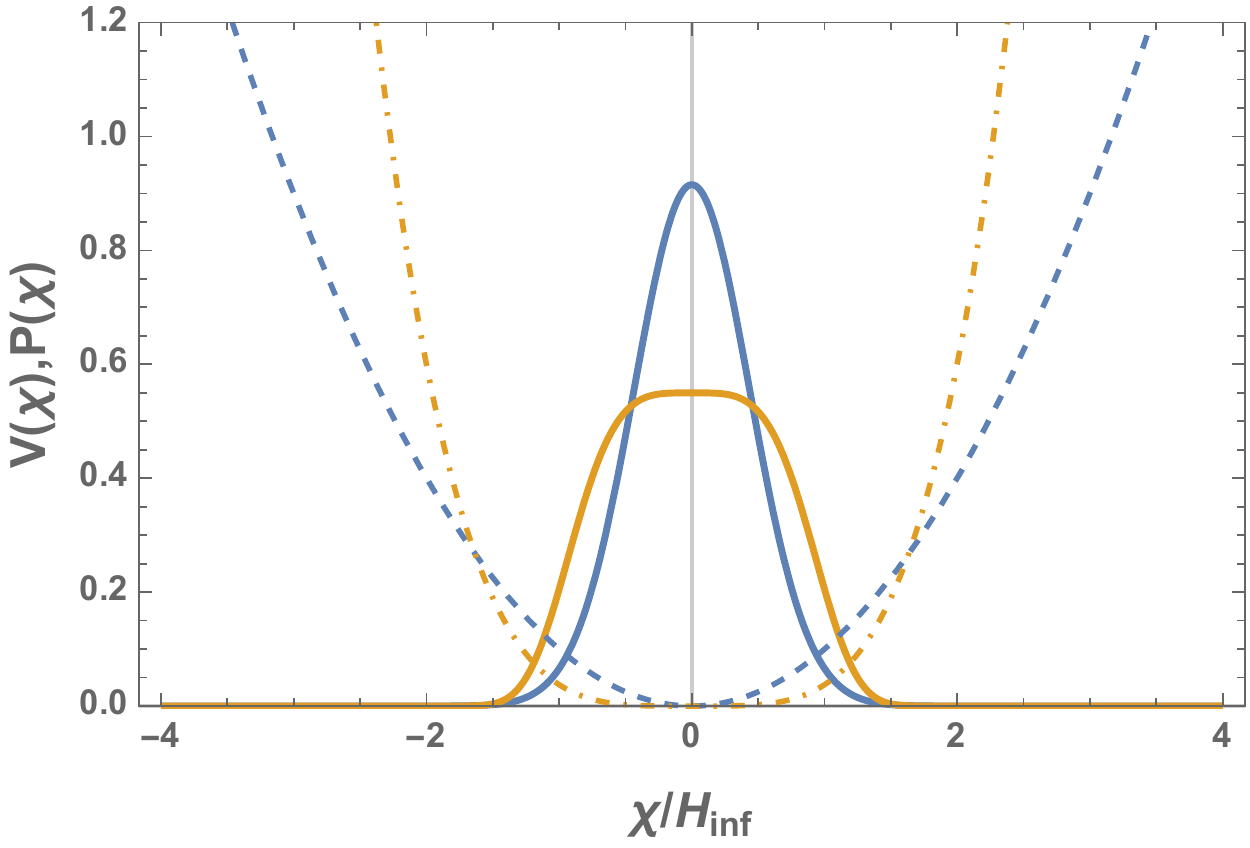}
\includegraphics[width=0.475\textwidth]{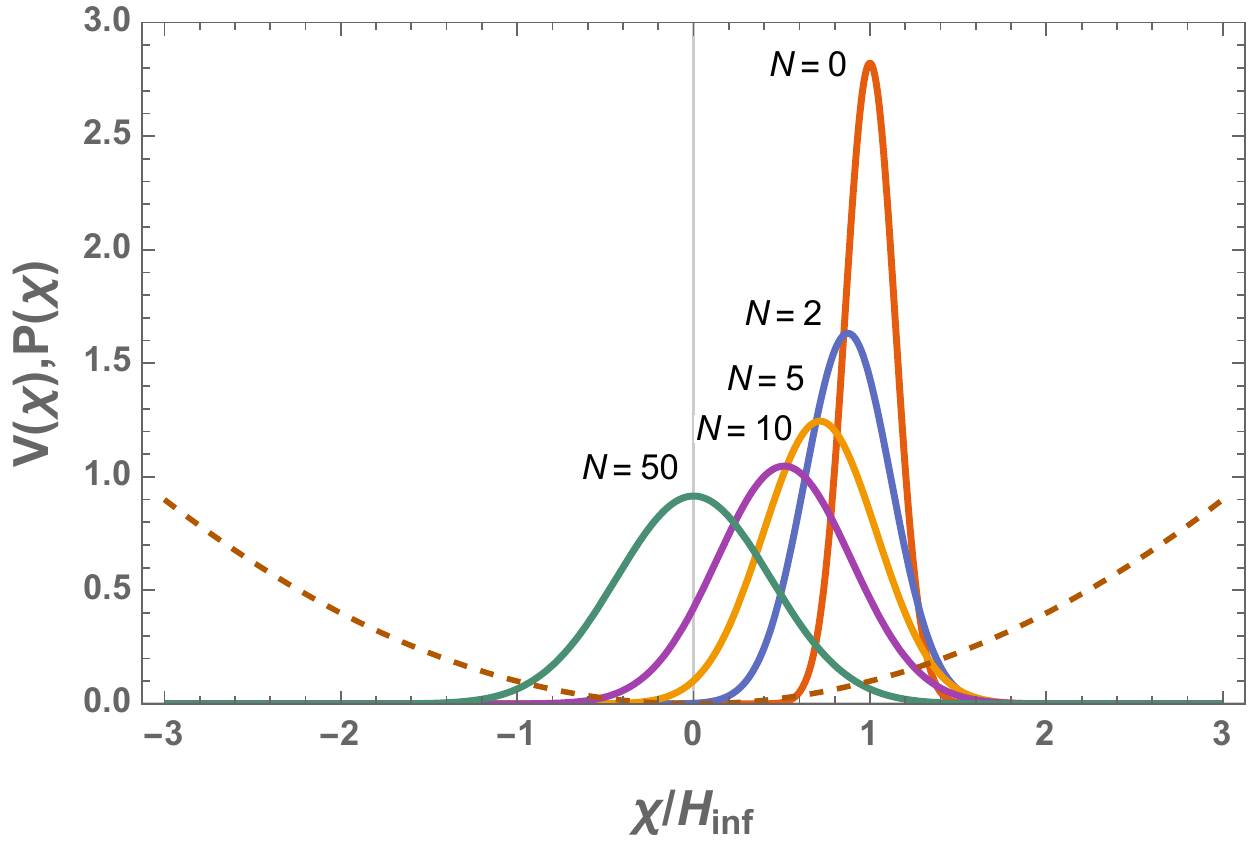}
\par\end{centering}
\caption{{\it Left panel}: Equilibrium distributions $P(\chi)$ in the quartic and quadratic cases (blue and orange thick curves, respectively) with the respective potentials shown in the background (blue dashed and orange dot-dashed curves, respectively). In this figure $m^2=0.2H_{\rm inf}^2$ and $\lambda = 0.1$. {\it Right panel}: Example of the relaxation of the (unnormalized) distribution function during inflation, here assuming a narrow Gaussian initial distribution and a quadratic potential with $m^2=0.2H_{\rm inf}^2$ (shown by the dashed curve). Shown to the left of each curve is the number of e-folds elapsed since the initial state. In roughly $50$ e-folds, the distribution reaches the equilibrium state given by Eq. \eqref{eq:p}.
 }
\end{figure}

Assuming that the effective mass of the spectator field $\chi$ was smaller than the Hubble rate during inflation, $V''<9H_{\rm inf}^2/4$ where the prime denotes derivative with respect to the field, it received quantum fluctuations from the rapidly expanding background. Using the stochastic approach \cite{Starobinsky:1986fx,Starobinsky:1994bd} (see also Refs. \cite{Kunimitsu:2012xx,Tokuda:2017fdh,Hardwick:2018sck,Tokuda:2018eqs,Hardwick:2019uex,Markkanen:2019kpv,Gorbenko:2019rza,Mirbabayi:2019qtx,Moreau:2019jpn,Baumgart:2019clc,Adshead:2020ijf,Markkanen:2020bfc,Bounakis:2020jdx,Moreau:2020gib,Pinol:2020cdp} for recent works), we find that the long wavelength modes of the field evolve according to the Langevin equation
\begin{equation}
\dot{\chi}(\bar{x},t) + \frac{1}{3H_{\rm inf}}V'(\chi) = f(\bar{x},t)\,,
\end{equation}
where $f(\bar{x},t)$ is a Gaussian noise term with
\begin{equation}
\langle f(\bar{x}_1,t_1)f(\bar{x}_2,t_2)\rangle = \frac{H_{\rm inf}^3}{4\pi^2}\delta(t_1-t_2)\,,
\end{equation}
and the point $\bar{x}$ is to be understood as a patch slightly larger than the Hubble volume during inflation, i.e. the field is coarse-grained over the Hubble horizon. Using standard techniques, one can turn the Langevin equation for the field into a Fokker-Planck equation for the one-point probability distribution $P(\chi(\bar{x},t))$, which reads
\begin{equation}
\frac{\partial P(\chi(\bar{x},t))}{\partial t} = D_\chi P(\chi(\bar{x},t))\,,
\end{equation}
where $D_\chi$ is the differential operator 
\begin{equation}
D_\chi \equiv \frac{V''(\chi)}{3H_{\rm inf}} + \frac{V'(\chi)}{3H_{\rm inf}}\frac{\partial}{\partial\chi} + \frac{H_{\rm inf}^3}{8\pi^2}\frac{\partial^2}{\partial\chi^2}\,. 
\end{equation}
One can show that there is an equilibrium solution for the one-point distribution function, which is given by 
\begin{equation}
\label{eq:p}
P(\chi) =C \exp\left(-\frac{8\pi^2}{3H_{\rm inf}^4}V(\chi)\right)\,,
\end{equation}
where $C$ is a normalization factor ensuring total probability of unity. This distribution describes the ensemble of field values in patches the size of a region slightly larger than the Hubble horizon at the end of inflation, see Fig. \ref{regions}. Notably, the scalar field reaches this ``equilibrium state" in a characteristic time scale regardless of its initial distribution, after which the distribution does not evolve anymore. This is depicted in Fig. \ref{distributions}. The relaxation time scale in terms of inflationary e-folds $N\equiv \ln(a/a_0)$, where $a_0$ is the scale factor at some reference time during inflation when the field had the value $\chi_0$ over a Hubble volume, is \cite{Enqvist:2012xn}
\begin{equation}
\label{Nrel}
N_{\rm rel} \simeq 
\begin{cases}
\displaystyle 11.3/\sqrt{\lambda} &\quad \lambda\chi^2 \gg m^2 
		,\\   
		\displaystyle 3H_{\rm inf}^2/m^2 & \quad \lambda\chi^2 \ll m^2\,,
\end{cases}
\end{equation}
depending on which term in Eq. \eqref{potential} dominates the spectator potential during inflation.

Therefore, by assuming inflation lasted for long enough for the field to reach the equilibrium state\footnote{As can be seen from Eq. \eqref{Nrel}, for large enough $\lambda$ or $m/H_{\rm inf}$ the equilibrium state can be reached even within the final $\sim 60$ e-folds of inflation, i.e. between the time when the mode corresponding to our currently observable Universe exited the horizon and the end of inflation. After the equilibrium is reached, all information of the initial conditions has been erased. For scenarios where the final distribution of field values carries information of the initial state, see Refs. \cite{Hardwick:2017fjo,Torrado:2017qtr}.}, the typical $\chi$ value at the end of inflation is given by the variance of \eqref{eq:p} as
\begin{equation}
	\label{variance}
\langle{\chi}_{\rm end}^2\rangle = 
\begin{cases} 
\displaystyle\sqrt{\frac{3}{2\pi^2}}{\frac{\Gamma(\frac{3}{4})}
		{\Gamma(\frac{1}{4})}} \frac{H_{\rm inf}^2}{\sqrt{\lambda}}\approx 0.132 \frac{H_{\rm inf}^2}{\sqrt{\lambda}}\,,& \quad \lambda\chi^2 \gg m^2 \,,
		\\   
		\displaystyle \frac{3}{8\pi^2}\frac{H_{\rm inf}^4}{m^2}\,, & \quad \lambda\chi^2 \ll m^2\,.
		\end{cases}
		\end{equation}
It should be noted, however, that while the variance $\langle{\chi}_{\rm end}^2\rangle $ describes the {\it typical} field value, there can be large variations in the field value in different Hubble patches at the time of photon decoupling. These constitute potentially dangerous DM isocurvature perturbations, which provide the most stringent constraints on our scenario (and also, on the flip side, the best potential for testability, as we will discuss). Before discussing the scalar field's perturbation spectrum, however, we will discuss the post-inflationary dynamics of the field and the dark matter production.

%%%%%%%%%%%%%%%%%%%%%%%%%%%%%%%%%%%%%%%%%%%%%%%%%%%%%%%%%%%%%%%%%%%%%%%%%%%%%%%%%%%%%%%%%%%%%%%

\section{Dynamics after inflation}
\label{dynamics}

\subsection{Background dynamics}
\label{bg_dyn}

As Eq. \eqref{variance} shows, in a typical situation the field is displaced away from its potential and has a finite initial value $\chi_{\rm end}\equiv \chi(\bar{x},t_{\rm end})$ over a patch slightly larger than the size of the Hubble horizon at the end of inflation, where $t_{\rm end}$ denotes the end of inflation. The equation of motion for the field describing its post-inflationary dynamics thus is\footnote{Here we neglect the gradient term, as the length scale over which the field values are correlated is typically much larger than the Hubble horizon at the end of inflation \cite{Kunimitsu:2012xx}.}
\begin{equation}
\label{eom}
\ddot{\chi} + 3H(t)\dot{\chi} + V'(\chi) = 0,
\end{equation}
where the dots denote derivatives with respect to cosmic time $t$, and
\begin{equation}
\label{Hubble}
H(t) = \frac{H_{\rm inf}}{\left(1+\frac{3(1+w)}{2}H_{\rm inf}t\right)} \simeq \frac{2}{3(1+w)}\frac{1}{t}\,,
\end{equation}
where the latter result applies shortly after inflation. Here $w\equiv p/\rho$ is the (time-averaged) equation of state parameter for the background with the energy density $\rho$ and pressure $p$. In the following, we will consider three cases: the usual radiation-dominated ($w=1/3$, $\rho\propto a^{-4}$), matter-dominated ($w=0$, $\rho\propto a^{-3}$) and kination-dominated ($w=1$, $\rho\propto a^{-6}$) Universe, so that after inflation
\begin{equation}
\label{Hscaling}
H \propto 
\begin{cases}
\displaystyle a^{-3/2}, & \quad w=0, \\
\displaystyle a^{-2}, & \quad w=1/3, \\
\displaystyle a^{-3}, & \quad w=1\,.
\end{cases}
\end{equation}
If reheating is prompt, the Universe quickly becomes radiation dominated, $w=1/3$. On the other hand, an early matter-dominated epoch could arise due to e.g. slow post-inflationary reheating or massive metastable particles that began to dominate the total energy density at some early stage prior to Big Bang Nucleosynthesis (BBN); see Ref. \cite{Allahverdi:2020bys} for a recent review. Finally, scenarios with $1/3<w<1$ are encountered in models where the total energy density of the Universe is dominated by the kinetic energy of a scalar field, either through oscillations in a steep potential, e.g. $V(\phi) \propto \phi^p$ with $p>4$, or by an abrupt drop in the scalar potential in the direction of this field \cite{Turner}. The latter possibility is exactly what happens in e.g. the case of quintessential inflation \cite{Peebles:1998qn}, where the inflaton field makes a transition from potential energy domination to kinetic energy domination at the end of inflation, reaching values of $w$ close to unity. The bound $w\leq 1$ comes from the requirement that the sound speed of the dominant fluid does not exceed the speed of light. Potentially more exotic scenarios that change the expansion history of the Universe compared to the standard radiation-dominated case could also be realized, see e.g. Refs. \cite{Kasuya:2001pr,Boyle:2001du,Bento:2002ps,Cline:1999ts,Takahashi:2020car}. Here, however, we only consider the three more conservative cases above.

\subsection{Quadratic case}
\label{sec:quadratic}

Let us begin by discussing the simplest possible case where the bare mass term dominates the spectator potential both during and after inflation, $m^2 \gg \lambda\chi_{\rm end}^2$. The solution to the equation of motion \eqref{eom} is then given by
\begin{equation}
\label{phi_solution}
\chi(t) = \chi_{\rm end}\times
\begin{cases}
\displaystyle \frac{{\rm sin}(mt)}{mt}, & \quad w=0, \\
\displaystyle 2^{1/4}\Gamma\left(\frac{5}{4}\right)\frac{J_{1/4}(mt)}{\left(mt\right)^{1/4}}, & \quad w=1/3, \\
\displaystyle J_{0}(mt), & \quad w=1,
\end{cases}
\end{equation}
where $J_\nu$ is the Bessel function of rank $\nu$. This result agrees well with the usual assumption that the field starts to oscillate roughly when $H(t)\simeq m$. At late times, $mt\gg 1$, the solutions \eqref{phi_solution} oscillate rapidly with an amplitude
\begin{equation}
\label{chi0}
\chi_0(t) \propto
\begin{cases}
\displaystyle (mt)^{-1}, & \quad w=0, \\
\displaystyle (mt)^{-3/4}, & \quad w=1/3, \\
\displaystyle (mt)^{-1/2} & \quad w=1,
\end{cases}
\end{equation}
and therefore in all cases the field has the associated energy density
\begin{equation}
\label{rho_chi_scaling}
\rho_\chi = \frac12m^2\chi_0^2 \propto a^{-3} ,
\end{equation}
as can be verified by inspection of Eqs. \eqref{Hubble}, \eqref{Hscaling} and \eqref{chi0}. Therefore, regardless of the background scaling, in this case the $\chi$ field constitutes an effective cold dark matter (CDM) component from the moment it starts oscillating.

\begin{figure}[H]
\begin{centering}
\includegraphics[scale=0.55]{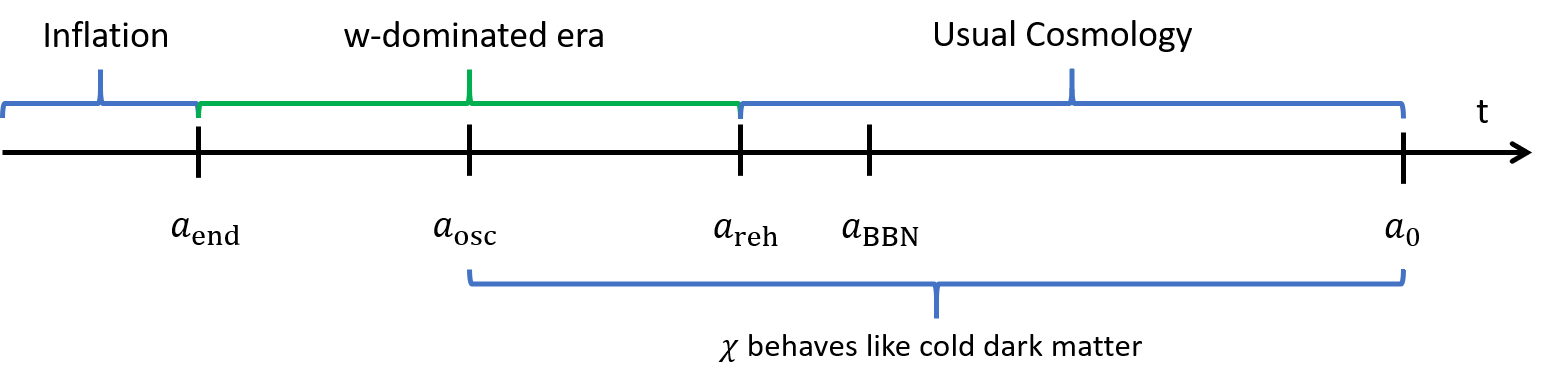}
\par\end{centering}
\caption{Cosmological evolution of the Universe in our model, in the case where the bare mass term dominates the spectator potential both during and after inflation.}
\label{history_quad}
\end{figure}

The cosmic history of our model is depicted in Fig. \ref{history_quad}.
As discussed above, the amplitude of the field remains frozen from the end of inflation
(which we denote by $a_{\mathrm{end}}$) roughly until its mass exceeds the Hubble parameter at $a_{\mathrm{osc}}$. At this moment, the field starts to oscillate and constitutes a DM component. At $a_{\mathrm{reh}}$, the $w$-dominated phase ends and the usual radiation-dominated era takes over the evolution of the Universe. In all cases considered in this paper, we assume that the dominant energy component causing the non-standard era decays only into radiation and does not affect the DM yield.

Thus, we can write the late time energy density of the field as 
\begin{eqnarray}
\label{chiscaling}
\rho_\chi(a) &=&  \rho_\chi(a_{\rm end})\left(\frac{a_{\rm osc}}{a_{\rm reh}}\right)^3\left(\frac{a_{\rm reh}}{a}\right)^3 \\ \nonumber
&=&\rho_\chi(a_{\rm end})\left(\frac{a_{\rm osc}}{a_{\rm end}}\right)^3\left(\frac{a_{\rm end}}{a_{\rm reh}}\right)^3\left(\frac{a_{\rm reh}}{a}\right)^3\,,
\end{eqnarray}
where the latter form is easier to evaluate, as
\begin{equation}
\left(\frac{a_{\rm osc}}{a_{\rm end}}\right)^3 = k^{\frac{4}{3(1+w)}}\left(\frac{H_{\rm inf}}{m}\right)^{\frac{2}{1+w}}\,,
\label{aosc aend}
\end{equation}
where $k\simeq 2.1$ is a factor that accounts for the fact that the oscillations do not start exactly when $H(t)=m$ and which we have evaluated by solving Eq. \eqref{eom} numerically,
\begin{equation}
\left(\frac{a_{\rm end}}{a_{\rm reh}}\right)^3 = \left(\frac{\rho_{\rm reh}}{\rho_{\rm end}}\right)^{\frac{1}{1+w}} = \left(\frac{\pi^2 g_*(T_{\rm reh})}{90}\right)^{\frac{1}{1+w}} \left(\frac{T_{\rm reh}^2}{H_{\rm inf}M_{\rm P}}\right)^{\frac{2}{1+w}}\,,
\label{aend areh}
\end{equation}
where $\rho_{\rm end}$ and $\rho_{\rm reh}$ correspond to the total energy density of the Universe at the given times, $T_{\rm reh}$ is the radiation temperature at the time when the non-standard expansion phase ended\footnote{We assume that the decay of the dominant energy density component causing the $w$-dominated era and the subsequent thermalization of SM particles were instantaneous, so that $\rho_{\rm reh}=\pi^2/30g_*T_{\rm reh}^4$. These are both fairly safe assumptions, as earlier studies have found that in terms of scaling of the background energy density, the transition from the $w$-domination to the usual radiation domination is very quick \cite{Berlin:2016gtr,Tenkanen:2016jic,Bernal:2018kcw} and once the dominant energy density component has decayed, the SM particles generically thermalize and build up a heat bath in much less than one e-fold from their production \cite{McDonough:2020tqq}.} and $g_*$ is the corresponding effective number of degrees of freedom, and $M_{\rm P}$ is the reduced Planck mass; and 
\begin{equation}
\label{areh_a}
\left(\frac{a_{\rm reh}}{a}\right)^3 = \frac{g_{*S}(T)}{g_{*S}(T_{\rm reh})}\left(\frac{T}{T_{\rm reh}}\right)^3\,,
\end{equation}
which follows from the fact that entropy is conserved after reheating. In all cases, we assume that the oscillations began before reheating, $a_{\rm osc}<a_{\rm reh}$, which amounts to requiring
\begin{equation}
\label{masscondition}
m > k^{\frac{2}{3}}\sqrt{\frac{\pi^2g_*(T_{\rm reh})}{90}}\frac{T^2_{\rm reh}}{M_{\rm P}}\,,
\end{equation}
independently of the inflationary scale $H_{\rm inf}$.

Thus, by substituting Eqs. \eqref{aosc aend}, \eqref{aend areh}, and \eqref{areh_a} into Eq. \eqref{chiscaling}, we find the present-day DM energy density
\begin{equation}
\label{DMabundance}
\Omega_\chi h^2 = \frac{k^{\frac{4}{3(1+w)}}}{2}\frac{g_{*S}(T_0)}{g_{*S}(T_{\rm reh})}\left(\frac{\pi^2g_*(T_{\rm reh})}{90}\right)^{\frac{1}{1+w}}\left(\frac{T_{\rm reh}^2}{mM_{\rm P}}\right)^{\frac{2}{1+w}}\left(\frac{T_0}{T_{\rm reh}}\right)^3\frac{m^2\chi_{\rm end}^2}{\rho_c/h^2}\,,
\end{equation}
where $T_0=2.725$ K is the present-day CMB temperature and $\rho_c/h^2 = 8.09\times 10^{-47}\, {\rm GeV}^4$ is the critical density. For suitable choices of the parameters $m\,,w\,,T_{\rm reh}$ and $\chi_{\rm end}$, the field can constitute all of the observed DM abundance, $\Omega_\chi h^2 =0.12$ \cite{Akrami:2018odb}, which in this case is produced by random fluctuations of the $\chi$ field during cosmic inflation. Note that while the local DM density is seemingly independent of the inflationary scale $H_{\rm inf}$, the typical field value (and therefore the typical DM density) is given by the variance of the field's fluctuation distribution, Eq. \eqref{variance}, which is determined by $H_{\rm inf}$. We can therefore use that equation for $\chi_{\rm end}^2$ to find the typical DM density. However, as we will show in Sec. \ref{perturbations}, maintaining the local field value $\chi_{\rm end}$ in \eqref{DMabundance} is crucial for determining the DM perturbation spectrum and therefore also in assessing the viability of the model. Finally, we note that by setting $w=1/3$, we find the result first obtained in Ref. \cite{Tenkanen:2019aij} modulo a factor 4 which was missing from Ref. \cite{Tenkanen:2019aij} but which has now been included. 

\subsection{Quartic case}
\label{sec:quartic}

Let us then consider the case where both during and right after inflation the scalar field's potential was dominated by the quartic term
\begin{equation}
V\left(\chi\right)\simeq\frac{\lambda}{4}\,\chi^{4}\,,
\end{equation}
and $m^2 \ll \lambda \chi_{\rm end}^2$. As shown in Appendix \ref{appendix}, in this case the scalar field equation of motion \eqref{eom} can be expressed in terms of conformal time ${\rm d}\eta ={\rm d}t/a$ as
\begin{equation}
\label{rescaled_eom}
z'' + F(\eta, w)z + z^3 = 0\,,
\end{equation}
where $z\equiv a\sqrt{\lambda}\chi$ is the rescaled field and $F(\eta, w)$ is given by Eq. \eqref{Fterm}. As shown in the Appendix, when $w=1/3$, Eq. (\ref{rescaled_eom}) reduces to
\begin{equation}
\label{quartic_RD_eom}
z'' + z^3 = 0\,,
\end{equation}
whose solution is a well-known oscillating function: the elliptic (Jacobi) cosine function, whose exact form can be found analytically (see e.g. Refs. \cite{Ichikawa:2008ne,Kainulainen:2016vzv}) and which, besides the oscillations, has no further time-dependence in terms of $\eta$. Because $\chi \propto z/a$, this means that when the background energy density is radiation-dominated, the oscillation amplitude decays simply as $\chi \propto 1/a$, and the spectator field behaves as dark radiation. Furthermore, as discussed in Appendix \ref{appendix}, the $F$-term in Eq. \eqref{rescaled_eom} dies off very quickly regardless of $w$, and the spectator field's equation of motion always reduces to \eqref{quartic_RD_eom}. Thus, in all cases we retain the usual $\chi \propto 1/a$ scaling, which validates the following treatment of the scalar field energy density.

\subsubsection{Coherent oscillations}

\begin{figure}[H]
\begin{centering}
\includegraphics[scale=0.55]{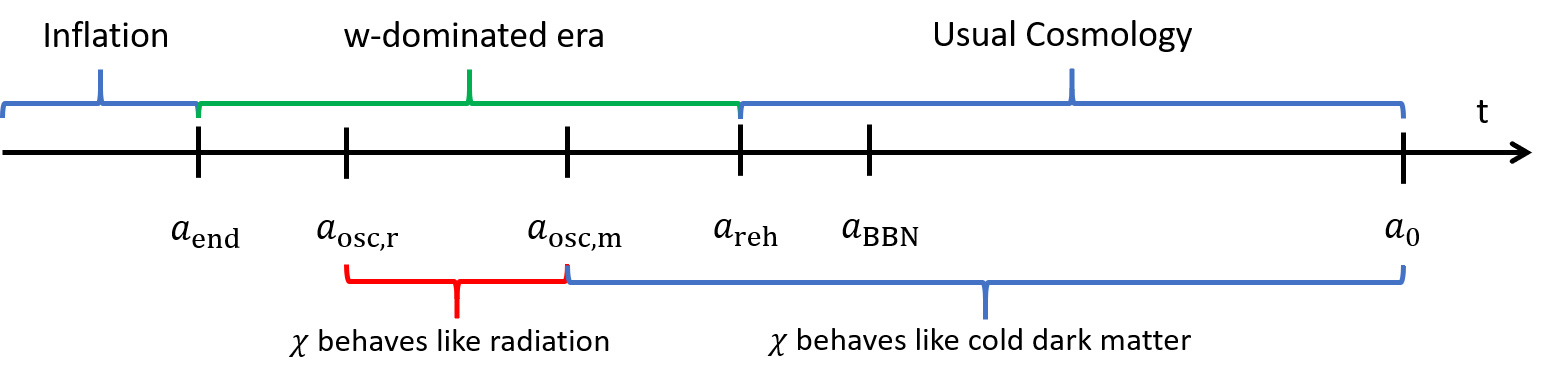}
\includegraphics[scale=0.55]{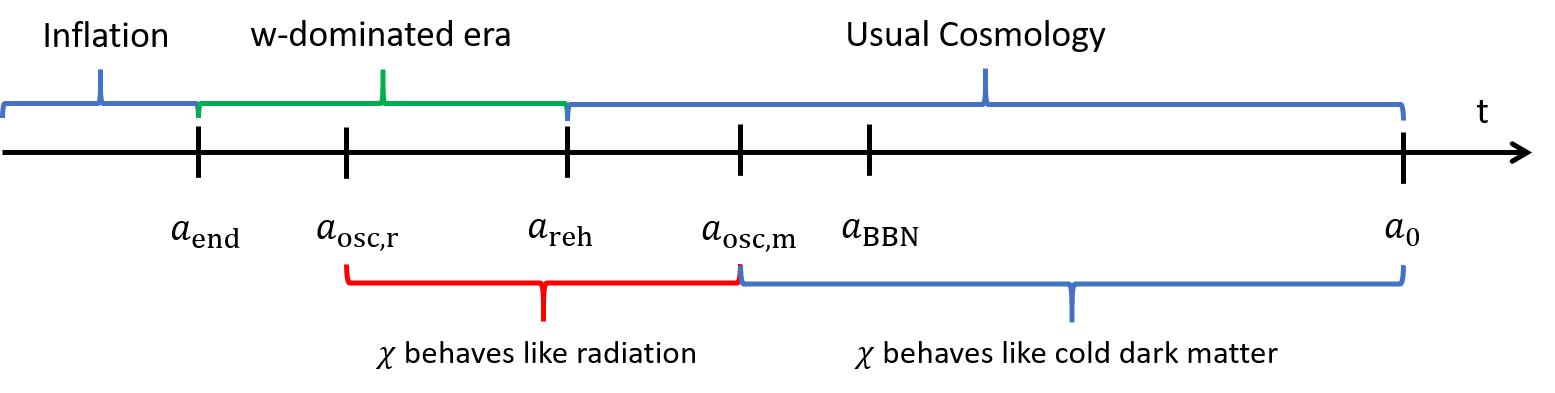}
\par\end{centering}
\caption{{\it Upper panel}: Cosmological evolution of the Universe in a scenario where the $\chi$ field behaves like dark radiation at early times after inflation and starts to behave like CDM before reheating.
{\it Lower panel}: Same as above but in this scenario, reheating occurs before the field starts to behave like CDM. The final DM abundance is the same in both cases.
 \label{fig:history}}
\end{figure}

As in the quadratic case discussed in Sec. \ref{sec:quadratic}, the field is in an overdamped regime roughly until its effective mass $V''(\chi) = 3\lambda\chi^{2}$ exceeds the Hubble parameter, after which $\chi$ starts to oscillate about its origin. We denote this moment by $a_{\mathrm{osc,r}}$, as the amplitude of the scalar field decays with $a^{-1}$ and the field constitutes a dark radiation component. At a time which we denote by $a_{\mathrm{osc,m}}$, the oscillation amplitude has decreased enough so that the quadratic term of the potential starts to dominate over the quartic one, and the field starts to behave as cold dark matter. For simplicity, we use the standard approximation where the energy density of $\chi$ instantaneously changes from scaling as $\rho_{\chi}\propto a^{-4}$ to $\rho_{\chi}\propto a^{-3}$ as soon as the quadratic term dominates, and assume in this subsection that the scalar field oscillations remained coherent throughout the above phases. The remaining of the cosmological history proceeds as in the scenario we studied in Sec. \ref{sec:quadratic}: the background field that is dominating the evolution of the Universe
decays at $a_{\mathrm{reh}}$ and the Universe enters into the usual radiation-dominated era, followed by a period of matter domination until the late-time dark energy domination finally takes over. The cosmic history is illustrated in Fig. \ref{fig:history}.

Regardless of when the scalar field reaches the quadratic part of its potential (before or after reheating), the energy density of the field is given by
\begin{equation}
\rho_{\chi}\left(a\right)=\frac{\lambda}{4}\,\chi_{\mathrm{end}}^{4}\,\left(\frac{a_{\mathrm{osc,r}}}{a_{\mathrm{osc,m}}}\right)^{4}\,\left(\frac{a_{\mathrm{osc,m}}}{a_{\mathrm{reh}}}\right)^{3}\,\left(\frac{a_{\mathrm{reh}}}{a}\right)^{3}.\label{rho DM late times}
\end{equation}
If the field started to oscillate about the {\it quartic} part of its potential only after reheating, there is no difference in the DM abundance between this scenario and the usual radiation-dominated one studied in Ref. \cite{Markkanen:2018gcw}. Therefore, in the following we will assume reheating always happens after the field has reached the quartic part of its potential.

Let us proceed by finding an expression for the present-day energy density of the field. First, we have
\begin{equation}
\frac{a_{\mathrm{osc,r}}}{a_{\mathrm{osc,m}}}=\left(\frac{H_{\mathrm{osc,m}}}{H_{\mathrm{osc,r}}}\right)^{\frac{2}{3\left(1+w\right)}}\,,\label{relation osc r m-1}
\end{equation}
where $H_{\rm osc,r}^2 = 3\lambda \chi_{\rm end}^2$, i.e. the field starts to oscillate when it becomes effectively massive. At $a_{\mathrm{osc,m}}$, the quadratic term in the equation of motion becomes equal to the quartic term
\begin{equation}
3\lambda\chi^2\left(a_{\mathrm{osc,m}}\right)=\lambda\chi^2_{\rm end}\left(\frac{a_{\mathrm{osc,r}}}{a_{\mathrm{osc,m}}}\right)^2=m^{2}\,,
\end{equation}
which gives us a relation between the Hubble parameters:
\begin{equation}
\frac{H_{\mathrm{osc,m}}}{H_{\mathrm{osc,r}}}=\left(\frac{m}{\sqrt{3\lambda}\left|\chi_{\rm end}\right|}\right)^{\frac{3\left(1+w\right)}{2}}.\label{H osc m-1}
\end{equation}
The factor $a_{\mathrm{osc,m}}/a_{\mathrm{reh}}$ can be obtained recalling that 
\begin{equation}
H(a_{\mathrm{reh}})=\sqrt{\frac{\pi^{2}\,g_{*}\left(T_{\mathrm{reh}}\right)}{90}}\,\frac{T_{\mathrm{reh}}^{2}}{M_{\mathrm{P}}}\,,
\end{equation}
so that
\begin{equation}
\left(\frac{a_{\mathrm{osc,m}}}{a_{\mathrm{reh}}}\right)^{3}=\left(\frac{m}{\sqrt{\lambda}\left|\chi_{\rm end}\right|}\right)^{-3}\,\left(\sqrt{\frac{\pi^{2}\,g_{*}\left(T_{\mathrm{reh}}\right)}{270}}\,\frac{T_{\mathrm{reh}}^{2}}{\sqrt{\lambda}\,\left|\chi_{\rm end}\right|M_{\mathrm{P}}}\right)^{\frac{2}{1+w}}. \label{aosc areh quart}
\end{equation}
Finally, for the ratio $a_{\rm reh}/a$ we can again use entropy conservation, Eq. \eqref{areh_a}.

Hence, by using the relations above, we conclude that the present DM abundance is in this case given by
\begin{equation}
\label{DMabundance_quartic}
\Omega_\chi h^2 =\frac{\sqrt{\lambda}}{4}\frac{g_{{*S}}(T_{0})}{g_{{*S}}(T_{{\rm reh}})}\,\left(\sqrt{\frac{\pi^{2}\,g_{*}\left(T_{\mathrm{reh}}\right)}{270}}\,\frac{T_{\mathrm{reh}}^{2}}{\sqrt{\lambda}\,\left|\chi_{\mathrm{end}}\right|M_{\mathrm{P}}}\right)^{\frac{2}{1+w}}\left(\frac{T_{0}}{T_{{\rm reh}}}\right)^{3}\frac{m\left|\chi_{\rm end}\right|^3}{\rho_c/h^2}\,.
\end{equation}
While this result for the DM abundance applies regardless of when the scalar field reaches the quadratic part of its potential, in deriving this result we assumed that the oscillations in the quartic part always start before reheating, $a_{\mathrm{osc,r}}<a_{\mathrm{reh}}$. This gives a lower bound on the self-coupling
\begin{equation}
\label{lambdalimit}
\lambda > \frac{\pi^2g_*(T_{\rm reh})}{270}\frac{T_{\rm reh}^4}{M_{\rm P}^2\chi_{\rm end}^2}\,,
\end{equation}
which also has to be taken into account for consistency of the calculation.

\subsubsection{Condensate evaporation}
\label{fragmentation}

If its interactions are sufficiently suppressed, the scalar field
behaves as a long-lived oscillating condensate, never fragmenting or reaching thermal equilibrium. This is the scenario considered in the previous section. However, if the self-interaction coupling $\lambda$ is large enough, the $\chi$ condensate may fragment into $\chi$ particles which thermalize into a WIMP-like
DM candidate as discussed in Refs. \cite{Ichikawa:2008ne,Nurmi:2015ema,Kainulainen:2016vzv,Heikinheimo:2016yds,Enqvist:2017kzh,Cosme:2018nly,Markkanen:2018gcw}. The condition for a
complete decay of the condensate, for quartic self-interactions, is
given by
\begin{equation}
\frac{\Gamma\left(\chi\left(a_{\mathrm{dec}}\right)\right)}{H_{\mathrm{dec}}}\simeq 0.013\lambda\left(\frac{a_{\rm dec}}{a_{\rm osc,r}}\right)^{\frac12(1+3w)}=1,\label{evaporation}
\end{equation}
where $\Gamma\left(\chi\left(a\right)\right) = 0.023\lambda^{3/2}\chi(a)$ is the
decay rate of the condensate into two $\chi$ particles \cite{Kainulainen:2016vzv}, $H_{\mathrm{dec}}$ is the Hubble parameter at the time of the decay at $a_{\mathrm{dec}}$ and the amplitude of the scalar field is $\chi\left(a\right)=\chi_{\rm end}\,\left(a_{\rm osc,r}/a\right)$. For simplicity, we assume that the decay always occurs prior to reheating, which amounts to requiring
\begin{equation}
\label{adec<areh}
\frac{a_{\rm dec}}{a_{\rm reh}} = (0.013\lambda)^{-\frac{2}{1+3w}}\left(\sqrt{\frac{\pi^2g_*(T_{\rm reh})}{270}}\frac{T_{\rm reh}^2}{\sqrt{\lambda}|\chi_{\rm end}|M_{\rm P}} \right)^{\frac{2}{3(1+w)}} < 1\,.
\end{equation}
The condensate can only fragment while in the quartic part of its potential \cite{Ichikawa:2008ne,Nurmi:2015ema,Kainulainen:2016vzv}, which in addition to Eq. \eqref{adec<areh} imposes an upper limit on the bare mass:
\begin{equation}
m^2< 3\lambda\chi^2_{\rm dec} = 3\lambda\,(0.013\lambda)^{\frac{4}{(1+3w)}}\chi^2_{\rm end}\,.
\label{bare mass lim}
\end{equation}
If the bare mass is larger than the limit (\ref{bare mass lim}), the condensate never fragments but remains oscillating until the present day, and the resulting DM abundance is given by Eq. \eqref{DMabundance_quartic}. In contrast, if the condition (\ref{bare mass lim}) is satisfied, the condensate fragments, the $\chi$ sector thermalizes with itself, and we need to compute the dark matter abundance from the freeze-out of $\chi$ particles from their internal thermal bath. 

In the following, we assume that the $\chi$ particles freeze out while still relativistic. The temperature the $\chi$ particles acquire after thermalization is obtained by equating the $\chi$ condensate's energy density to the usual form of radiation energy density, which gives
\begin{equation}
T_{\chi}\left(a\right)=\left(\frac{15\lambda}{2\pi^{2}}\right)^{1/4}\left|\chi_{\mathrm{end}}\right|\left(\frac{a_{\mathrm{osc,r}}}{a}\right)\,.
\label{chi T}
\end{equation}
Note that this temperature is different from the temperature of the SM particle heat bath and can also scale differently from it in terms of $a$, see e.g. Ref. \cite{Carlson:1992fn}. However, the $\chi$ particle number density corresponding to $T_\chi$ is given by the usual expression
\begin{equation}
n_{\chi}\left(a\right) =\frac{\zeta\left(3\right)}{\pi^{2}}\,T_{\chi}^{3}(a)\,.
\label{number density FO}
\end{equation}
After freeze-out, the $\chi$ particles do not interact anymore, which means that the above relation is valid even when $\chi$ becomes non-relativistic. The energy density of $\chi$ at the present time is therefore simply $\rho_{\chi}(a_{0}) =mn_{\chi}\left(a_{0}\right)$, leading to the
following present abundance:
\begin{equation}
\label{DM_abundance_FO}
\Omega_\chi h^2 =\left(\frac{15}{2}\right)^{3/4}\frac{\zeta(3)\lambda^{3/4}}{\pi^{7/2}}\frac{g_{{*S}}(T_{0})}{g_{{*S}}(T_{{\rm reh}})}\,\left(\sqrt{\frac{\pi^{2}\,g_{*}\left(T_{\mathrm{reh}}\right)}{270}}\,\frac{T_{\mathrm{reh}}^{2}}{\sqrt{\lambda}\,\left|\chi_{\mathrm{end}}\right|M_{\mathrm{P}}}\right)^{\frac{2}{1+w}}\left(\frac{T_{0}}{T_{{\rm reh}}}\right)^{3}\frac{m\left|\chi_{\rm end}\right|^3}{\rho_c/h^2}\,,
\end{equation}
where the main difference to the result in the case of coherent oscillations \eqref{DMabundance_quartic} is that thermalization of $\chi$ particles changes the result's dependence on $\lambda$.

Finally, we note that the DM freeze-out could also occur while the DM particles are non-relativistic. In this case, the scalar field
undergoes a phase of cannibalism \cite{Carlson:1992fn}, where the $4\rightarrow2$ self-annihilations dilute the number density of $\chi$ particles before their eventual freeze-out. While performing this calculation in the standard radiation-dominated case is relatively simple \cite{Markkanen:2018gcw}, in the presence of a non-standard epoch and entropy production this calculation becomes much more involved. While the qualitative picture is quite different from the case where DM freeze-out is determined by the $2\to 2$ scatterings (as above), the quantitative difference is very modest in the standard radiation-dominated case \cite{Markkanen:2018gcw} and it is expected to be small in non-standard cases as well; see Ref. \cite{Bernal:2018ins} for an example in a matter-dominated case. Therefore, in this paper we do not consider this possibility but leave it for future work.

\section{Dark matter perturbations}
\label{perturbations}

Because the $\chi$ field is assumed to be decoupled from the SM radiation, fluctuations in the local field value necessarily generate isocurvature perturbations between the DM and radiation energy densities. Due to the non-observation of isocurvature perturbations in the CMB, this provides the most stringent observational constraints on our scenario.

More precisely, the DM isocurvature perturbation is defined as
\begin{equation}
\label{isocurvature_def}
S_{r\chi} \equiv -3H\left(\frac{\delta\rho_r}{\dot{\rho_r}} - \frac{\delta\rho_\chi}{\dot{\rho_\chi}} \right),
\end{equation}
where $\rho_i$ is the energy density of the fluid $i=r,\chi$ and perturbations are defined as deviations from the average energy density of the fluid $i$,
\begin{equation}
\delta\rho_i \equiv \frac{\rho_i(x)}{\langle\rho_i\rangle} -1\,.
\end{equation}
As discussed in Sec. \ref{inflation}, we assume that the perturbations in radiation energy density were sourced by the inflaton field, whereas the perturbations in the DM energy density were also sourced by the spectator field. Because the fluids are assumed to be decoupled from each other, we obtain 
\begin{equation}
H\frac{\delta\rho_i}{\dot{\rho_i}} = \frac{\delta\rho_i}{3(1+w_i)\rho_i}\,, 
\end{equation}
where $w_i \equiv p_i/\rho_i$ is the effective equation of state parameter of the fluid $i$ which relates the pressure of the fluid to its energy density, i.e. $w_r=1/3$ for radiation and $w_\chi=0$ for the spectator field at late times. The isocurvature perturbation then becomes
\begin{equation}
\label{Srchi}
S_{r\chi}  = \frac{\delta f(\chi_{\rm end})}{\langle f(\chi_{\rm end})\rangle}\,,
\end{equation}
where in the quadratic case $f(\chi_{\rm end}) = \chi_{\rm end}^2$ and in the quartic case $f(\chi_{\rm end}) = |\chi_{\rm end}|^{3-2/(1+w)}$, as given in Eqs. \eqref{DMabundance} and \eqref{DMabundance_quartic}, \eqref{DM_abundance_FO}, respectively. Note that because the mean field value vanishes, $\langle \chi_{\rm end}\rangle = 0$, it would be incorrect to assume $\delta f(\chi_{\rm end}) \propto \delta \chi_{\rm end}/\chi_{\rm end}$.

The isocurvature perturbation spectrum can be found in terms of the stochastic correlation functions that describe the field fluctuations during inflation \cite{Starobinsky:1994bd,Markkanen:2019kpv}. One finds that the power spectrum of the equal-time correlator of an arbitrary function of the scalar field $f(\chi)$ is given by \cite{Markkanen:2019kpv}
\begin{equation}
\label{Pfk}
\mathcal{P}_f(k) =\mathcal{A}_f\left(\frac{k}{H_{\rm inf}}\right)^{n_f-1}\,, 
\end{equation}
where 
\begin{equation}
\label{Af}
\mathcal{A}_f =  \frac{2}{\pi}f_n^2\Gamma\left[2-(n_f-1)\right]\sin\left(\frac{\pi (n_f-1)}{2}\right) 
\end{equation}
and
\begin{equation}
\label{nf}
n_f - 1 = \frac{2\Lambda_n}{H_{\rm inf}}\,,
\end{equation}
which applies for all modes $k\ll H_{\rm inf}$, i.e. for physical distance scales much larger than the horizon at the end of inflation. Because the CMB measurements are made at scales which are exponentially larger than $H_{\rm inf}^{-1}$, the form of \eqref{Pfk} is indeed suitable for our purposes.

As discussed in Ref.~\cite{Markkanen:2019kpv} (see also Refs.~\cite{Markkanen:2018gcw,Tenkanen:2019cik}) the parameters $f_n$ and $\Lambda_n$ are related to the eigenfunctions and eigenvalues of the Schr\"odinger-like equation
\begin{equation}
\label{Schrodinger}
\left(\frac{1}{2}\frac{\partial^2}{\partial\chi^2}-\frac{1}{2}\left(v'(\chi)^2-v''(\chi)\right) \right) \psi_n(\chi)
=-\frac{4\pi^2\Lambda_n}{H_{\rm inf}^3}\psi_n(\chi)\,,
\end{equation}
where
\begin{equation}
    v(\chi)=\frac{4\pi^2}{3H_{\rm inf}^4}V(\chi) = 
   \begin{cases} 
    \displaystyle \frac{\pi^2\lambda}{3}\left(\frac{\chi}{H_{\rm inf}}\right)^4\,, &\quad \lambda\chi^2 \gg m^2\,,\\   
		    \displaystyle \frac{2\pi^2}{3}\left(\frac{m}{H_{\rm inf}}\right)^2\left(\frac{\chi}{H_{\rm inf}}\right)^2\,, &\quad \lambda\chi^2 \ll m^2\,,
    \end{cases}
\end{equation}
and $f_n$ is given in terms of the eigenfunctions in \eqref{Schrodinger} as
\begin{equation}
f_n=\int d\phi \psi_0(\chi)f(\chi)\psi_n(\chi)\label{eq:fn}\,.
\end{equation}
These quantities enter the calculation of the DM isocurvature spectrum through the spectral expansion of the unequal-time correlator\footnote{By using de Sitter invariance, this result can be used to find an expression for the equal-time correlator power spectrum \eqref{Pfk}. Here we present only the most important steps; for more details on the derivation of this result, see Refs. \cite{Starobinsky:1994bd,Markkanen:2019kpv}.}
\begin{equation}
\langle f(\chi(0))f(\chi(t))\rangle = \sum_n f_n^2 e^{-\Lambda_n t}\,,
\end{equation}
where only the first non-trivial term is important, as the higher-order corrections are exponentially suppressed \cite{Markkanen:2019kpv}. In the quadratic case, $\lambda\chi^2 \ll m^2$, we find analytically
\begin{equation}
\label{Lambda_f_quadr}
\Lambda_2^{(2)} = \frac23\frac{m^2}{H_{\rm inf}}\,, \quad f^{(2)}_2 = \sqrt{2}\,,
\end{equation}
where the superscripts denote the quadratic case, whereas in the quartic case, $\lambda\chi^2 \gg m^2$, a numerical solution of the eigenvalue equation (\ref{Schrodinger}) gives
\begin{equation}
\label{Lambda_quart}
    \Lambda_2^{(4)}\approx 0.289\sqrt{\lambda}H_{\rm inf}
\end{equation}
and
\begin{equation}
\label{f2_quart}
f_2^{(4)}(w)\approx 
\begin{cases}
0.639 & \quad w=0\,,\\
0.867 & \quad w=1/3\,,\\
1.057 & \quad w=1\,,\\
\end{cases}
\end{equation}
where we used the fact in the quartic case $f(\chi_{\rm end}) = |\chi_{\rm end}|^{3-2/(1+w)}$, as discussed below Eq. \eqref{Srchi}.

By substituting the results \eqref{Lambda_f_quadr}--\eqref{f2_quart} into Eqs. \eqref{Af} and \eqref{nf}, we find the DM isocurvature power spectrum as
\begin{equation}
\label{PS}
\mathcal{P}_S = \mathcal{A}_S\left(\frac{k}{k_*}\right)^{n_S-1}\,,
\end{equation}
where the amplitude at $k=k_*$ is given by
\begin{equation}
\label{S_amplitude}
\mathcal{A}_S =
\begin{cases}
\displaystyle \frac{2(f_2^{(4)}(w))^2}{\pi}\Gamma\left[2-(n_S-1)\right]\sin\left(\frac{\pi (n_S-1)}{2}\right)e^{-(n_S - 1)N(k_*)},  & \quad \lambda\chi^2 \gg m^2, \\
\displaystyle \frac{4}{\pi}\Gamma\left[2-(n_S-1)\right]\sin\left(\frac{\pi (n_S-1)}{2}\right) e^{-(n_S - 1)N(k_*)}, & \quad \lambda\chi^2 \ll m^2,
\end{cases}
\end{equation}
where $N(k_*)$ is the number of e-folds between the horizon exit of a scale $k_*$ and the end of inflation, and the spectral tilt is
\begin{equation}
\label{nS}
n_S - 1 = 
\begin{cases}
\displaystyle  0.579\sqrt{\lambda}, & \quad \lambda\chi^2 \gg m^2\,, \\
\displaystyle \frac{4}{3}\frac{m^2}{H_{\rm inf}^2}, & \quad \lambda\chi^2 \ll m^2\,.
\end{cases}
\end{equation}
As the reference (pivot) scale we use $k_* = 0.05\,{\rm Mpc}^{-1}$, which is also one of the pivot scales the Planck collaboration used in their analysis. Following their conventions, the result \eqref{PS} should be compared to the observational constraint for an uncorrelated DM isocurvature perturbation 
\begin{equation}
\label{beta}
\mathcal{P}_S(k_*) = \frac{\beta}{1-\beta}\mathcal{P}_\zeta(k_*)\,,
\end{equation}
where $\beta < 0.38$ and $\mathcal{P}_\zeta(k_*)= 2.1\times 10^{-9}$ is the observed amplitude of the curvature power spectrum \cite{Akrami:2018odb}.

To compute the DM perturbation spectrum in terms of our free parameters, we need to know the number of e-folds between horizon exit of the pivot scale and the end of inflation. As the result \eqref{S_amplitude} shows, the DM isocurvature power spectrum is exponentially sensitive to this number and therefore the differences between different scenarios can be large. In particular, this applies to deviations from the usual radiation-dominated case, which makes it interesting to consider such scenarios and in this way also constrain them.

\begin{figure}[httb]
\begin{centering}
\includegraphics[scale=.8]{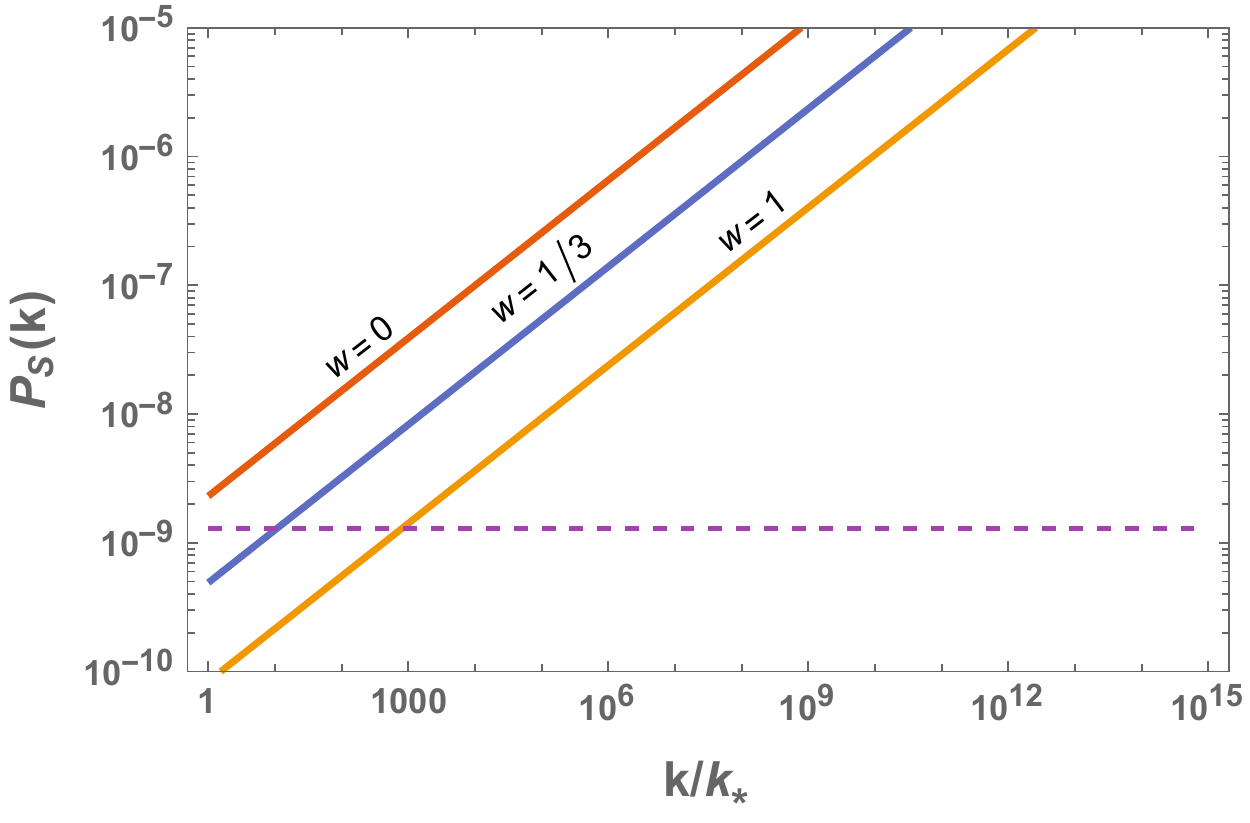}
\par\end{centering}
\caption{The DM isocurvature power spectrum $\mathcal{P}_S(k)$ as a function of $k/k_*$ in the quartic case in three different scenarios, from top to bottom: $w=0\,, 1/3\,, 1$. The horizontal dashed line shows the CMB isocurvature constraint $\mathcal{P}_S(k_*) < 1.3\times 10^{-9}$ at the CMB pivot scale $k_*=0.05\,{\rm Mpc}^{-1}$ \cite{Akrami:2018odb}. In this figure $H_{\rm inf}=10^{13}$ GeV, $T_{\rm reh}=10^8$ GeV, $\lambda = 0.4$.  The figure shows that for the above parameters, the case $w=0$ is not allowed by the CMB observations, whereas the cases with $w=1/3$ or $w=1$ are viable.}
 \label{fig:P_S}
\end{figure}

The number of e-folds between the horizon exit of a scale $k$ and the end of inflation is given by (see e.g. Ref. \cite{Allahverdi:2020bys})
\be
\label{Nk}
N(k) = \ln\left(\frac{a_{\rm end}}{a_{\rm reh}}\right) + \ln\left(\frac{a_{\rm reh}}{a_0}\right) + \ln\left(H_{\rm inf} k^{-1}\right)\,,
\ee
where the ratio of the scale factors at the end of inflation and at the time of reheating is
\begin{eqnarray}
\label{ae arh}
\ln \left(\frac{a_{\rm end}}{a_{\rm reh}}\right) &=& \frac{1}{3(1+w)}\ln\left(\frac{\rho_{\rm reh}}{\rho_{\rm end}}\right)  \\ \nonumber
&=& \frac{1}{3(1+w)}\left[\ln\left(\frac{\pi^2 g_*(T_{\rm reh})}{90}\right) + 4\ln\left(\frac{T_{\rm reh}}{M_{\rm P}}\right) - 2\ln\left(\frac{H_{\rm inf}}{M_{\rm P}}  \right) \right]\,.
\end{eqnarray}
Because the result depends only logarithmically on $g_*(T_{\rm reh})$ (and is further suppressed by the $w$-dependent prefactor), this quantity should differ from the usual value ($\sim 100$) by orders of magnitude in order to affect the result. Therefore, here we simply assume $g_*(T_{\rm reh})\sim 100$, which allows to write Eq. \eqref{ae arh} as
\begin{equation}
\ln \left(\frac{a_{\rm end}}{a_{\rm reh}}\right) \simeq \frac{1}{3(1+w)}\left[4\ln\left(\frac{T_{\rm reh}}{M_{\rm P}}\right) - 2\ln\left(\frac{H_{\rm inf}}{M_{\rm P}}  \right) \right]\,.
\end{equation}
We also have
\be
\ln\left(\frac{a_{\rm reh}}{a_0}\right) = \frac13\ln\left(\frac{g_{*S}(T_0)T_0^3}{g_*(T_{\rm reh})T^3_{\rm reh}} \right) \simeq -72.5 - \ln\left(\frac{T_{\rm reh}}{M_{\rm P}}\right),
\ee
where we used $g_{*S}=3.909$, $T_0=2.725$ K; and
\be
\ln\left(H_{\rm inf} k^{-1}\right) \simeq 133.3 + \ln\left(\frac{H_{\rm inf}}{M_{\rm P}}\right) - \ln\left(\frac{k}{0.05\,{\rm Mpc}^{-1}}\right)\,.
\ee
Thus, putting all of the above results together, we obtain for the e-fold number corresponding to the pivot scale $k_*=0.05\,{\rm Mpc}^{-1}$ the result
\be
\label{N pivot}
N(k_*) \simeq 60.8 + \frac{1}{3(1+w)}\left[4\ln\left(\frac{T_{\rm reh}}{M_{\rm P}}\right) - 2\ln\left(\frac{H_{\rm inf}}{M_{\rm P}}  \right) \right] + \ln\left(\frac{H_{\rm inf}}{M_{\rm P}}\right) - \ln\left(\frac{T_{\rm reh}}{M_{\rm P}}\right)\,.
\ee
It should be noted, however, that for the assumptions made in this paper (in particular about the Hubble parameter that stays roughly constant during inflation), the number of e-folds is bounded from above as $N(k_*)\lesssim 63$ due to the BBN constraints on gravitational waves\footnote{In the presence of a stiff era, $w>1/3$, gravitational waves become enhanced compared to the case with $w\leq 1/3$ and can contribute to the $N_{\rm eff}$ parameter in a significant way, allowing one to constrain such scenarios. For recent works, see e.g. Refs. \cite{Caprini:2018mtu,Bernal:2019lpc,Figueroa:2019paj}.} \cite{Tanin:2020qjw}, which affects our results in the $w=1$ case. In general, we will use the result \eqref{N pivot} to evaluate the DM isocurvature perturbation spectrum amplitude \eqref{S_amplitude}, which we will contrast with observations through Eq. \eqref{beta}. For an illustration of the resulting $P_S(k)$ in a few example cases, see Fig.~\ref{fig:P_S}. 

%%%%%%%%%%%%%%%%%%%%%%%%%%%%%%%%%%%%%%%%%%%%%%%%%%%%%%%%%%%%%%%%%%%%%%%%%%%%%%%%%%%%%%%%%%

\section{Results}
\label{results}

Finally, we present the requirements for the field $\chi$ to constitute all DM in the Universe. In addition to satisfying the DM abundance, there are a few requisites -- either observational constraints or consistency conditions specific to our scenario -- that constrain the model, and we begin by listing them here.

\subsection{Constraints on the scenario}
\label{sec:constraints}

First, the maximum Hubble scale during inflation is
\begin{equation}
\label{H_inf}
H_{\rm inf} =\sqrt{\frac{\pi^{2}rP_{\zeta}}{2}}M_{{\rm P}} \lesssim 6\times10^{13}\,\sqrt{\frac{r}{0.06}}\,,
\end{equation}
which follows from the definition of the tensor-to-scalar ratio $r$ and the usual slow-roll approximation, see e.g. Ref. \cite{Baumann:2009ds}. Here we have normalized $r$ to the largest value allowed by observations, $r\leq 0.06$ \cite{Akrami:2018odb}. In all our results, we assume that the field's fluctuation spectrum is not suppressed during inflation, $V''<9H_{\rm inf}^2/4$, which provides an upper limit on the parameter that characterizes the shape of the potential in each case, $\lambda$ or $m$, whereas the isocurvature constraint \eqref{beta} imposes additional constraints on them\footnote{For completeness, we note that there is also another branch of solutions to Eq. \eqref{beta}, which in the quartic case requires a very small value of the self-coupling, $\lambda\lesssim \mathcal{O}(10^{-19})$. As this regime is phenomenologically less interesting, here we neglect this possibility. For $m/H_{\rm inf}$ the solutions corresponding to the other branch are shown in Figs. \ref{w0 quad} and \ref{w1 quad}.}. The maximum reheating temperature for a given Hubble scale during inflation is
\begin{equation}
\label{TrehMax}
T_{{\rm reh}} \leq \left(\frac{90}{\pi^{2}g_{*}(T_{{\rm reh}})}\right)^{1/4}\sqrt{H_{{\rm inf}}M_{{\rm P}}}\,,
\end{equation}
which for the SM degrees of freedom, $g_*(T_{\rm reh}) = 106.75$, gives the absolute maximum reheating temperature as $T_{{\rm reh}}^{\rm max} = 6.7\times10^{15}\left(r/0.06\right)^{1/4}\,{\rm GeV}$, which follows from Eq. \eqref{H_inf}. However, the condition \eqref{TrehMax} is more general and should be applied for each $H_{\rm inf}$ separately. On the other hand, to not interfere with the formation of light elements, the non-standard phase has to end early enough so that the Universe gets reheated to a sufficiently high temperature. In the following, we will use the requirement $T_{\rm reh}\geq 10$ MeV to account for this aspect.

In the quartic case, observations of collisions between galaxy clusters (including the Bullet Cluster) can be used to place an upper bound on the self-interaction cross-section over DM mass, $\sigma/m\le 1$ cm$^2$/g $\approx 4.6\times 10^3\,{\rm GeV}^{-3}$ ~\cite{Markevitch:2003at,Randall:2007ph,Rocha:2012jg,Peter:2012jh,Harvey:2015hha}. For our theory \cite{Heikinheimo:2016yds}
\begin{equation}
    \frac{\sigma}{m} = \frac{9\lambda^2}{32\pi m^3} ,
\end{equation}
so the the galaxy collisions impose a constraint
\begin{equation}
\label{eq:sigmaDMbound}
\frac{m}{\rm GeV} > 0.027\left(\frac{\sigma/m}{{\rm cm}^2/{\rm g}}\right)^{-1/3} \lambda^{2/3}\,,
\end{equation}
which limits the quartic scenario at sub-GeV masses for detectable values of $\sigma/m$.

Finally, in deriving the results for DM abundance, Eq. \eqref{DMabundance} in the quadratic case and Eqs. \eqref{DMabundance_quartic} or \eqref{DM_abundance_FO} in the quartic case (depending on whether the condensate fragments or not), we made the assumption that the post-inflationary oscillations of the $\chi$ field always began prior to reheating\footnote{While relaxing this assumption does not change the results for the DM abundance found in Refs. \cite{Markkanen:2018gcw,Tenkanen:2019aij}, an early phase of non-standard expansion would change the isocurvature limit through the altered number of e-folds, Eq. \eqref{N pivot}. In this paper, however, we do not account for this possibility in our figures for simplicity.}, $a_{\rm osc}/a_{\rm reh}<1$, which amounts to requiring Eq. \eqref{masscondition} in the quadratic case and Eq. \eqref{lambdalimit} in the quartic case. We also assumed that the possible decay of the oscillating condensate always occurs prior to reheating, which imposes the conditions given by Eqs. \eqref{adec<areh} and \eqref{bare mass lim}. In addition to the observational constraints discussed above, these consistency conditions provide constraints on the model parameter space. They are all accounted for in the results shown in the next subsection.

\subsection{Model parameter space}

Next, we present the results by assuming in all cases that the DM abundance is given by the typical field value, $\chi_{\rm end}^2 = \langle \chi_{\rm end}^2 \rangle$, as in Eq. \eqref{variance}. 

\begin{figure}[H]
\begin{centering}
\includegraphics[width=0.48\textwidth]{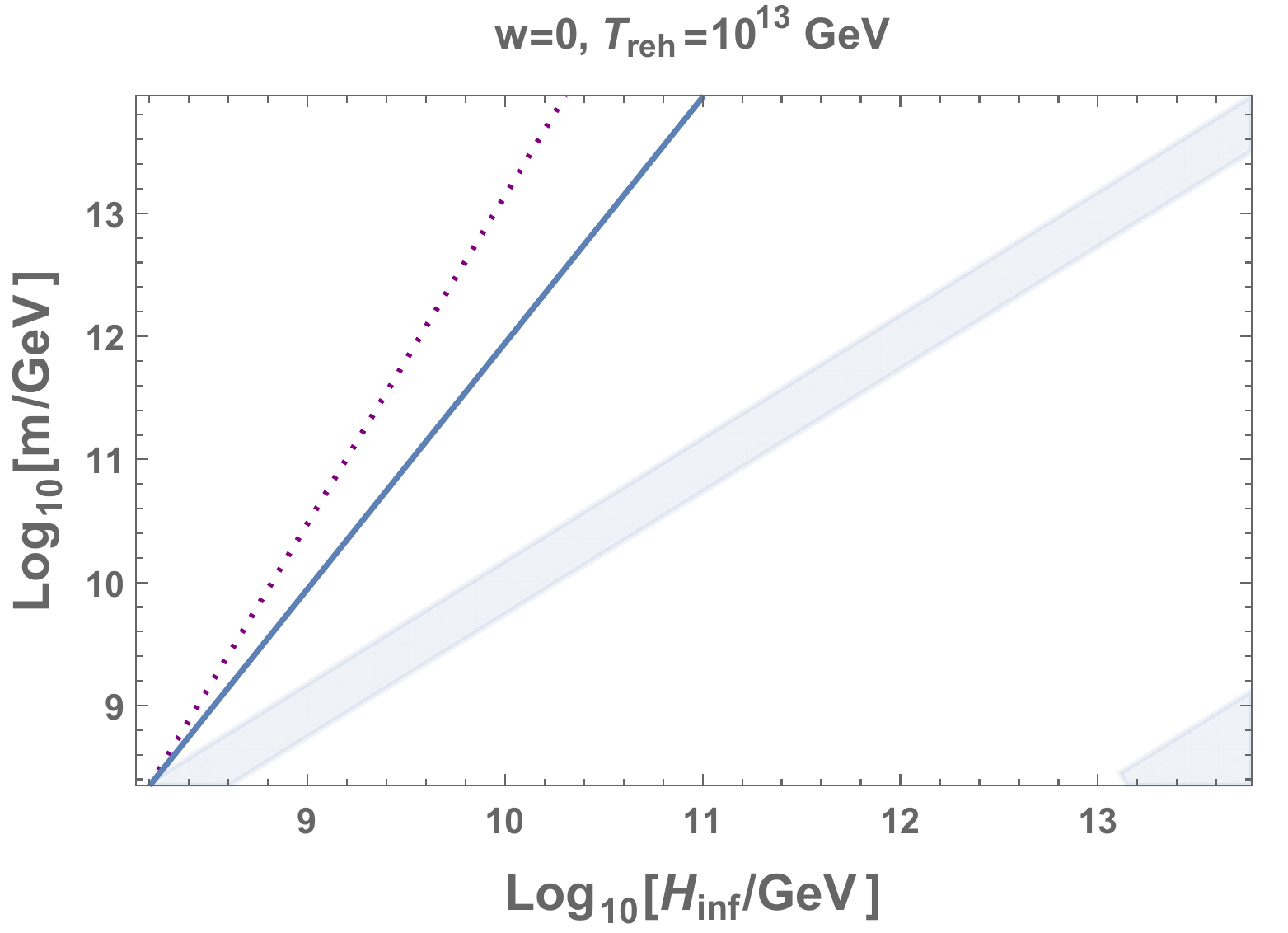}
\includegraphics[width=0.48\textwidth]{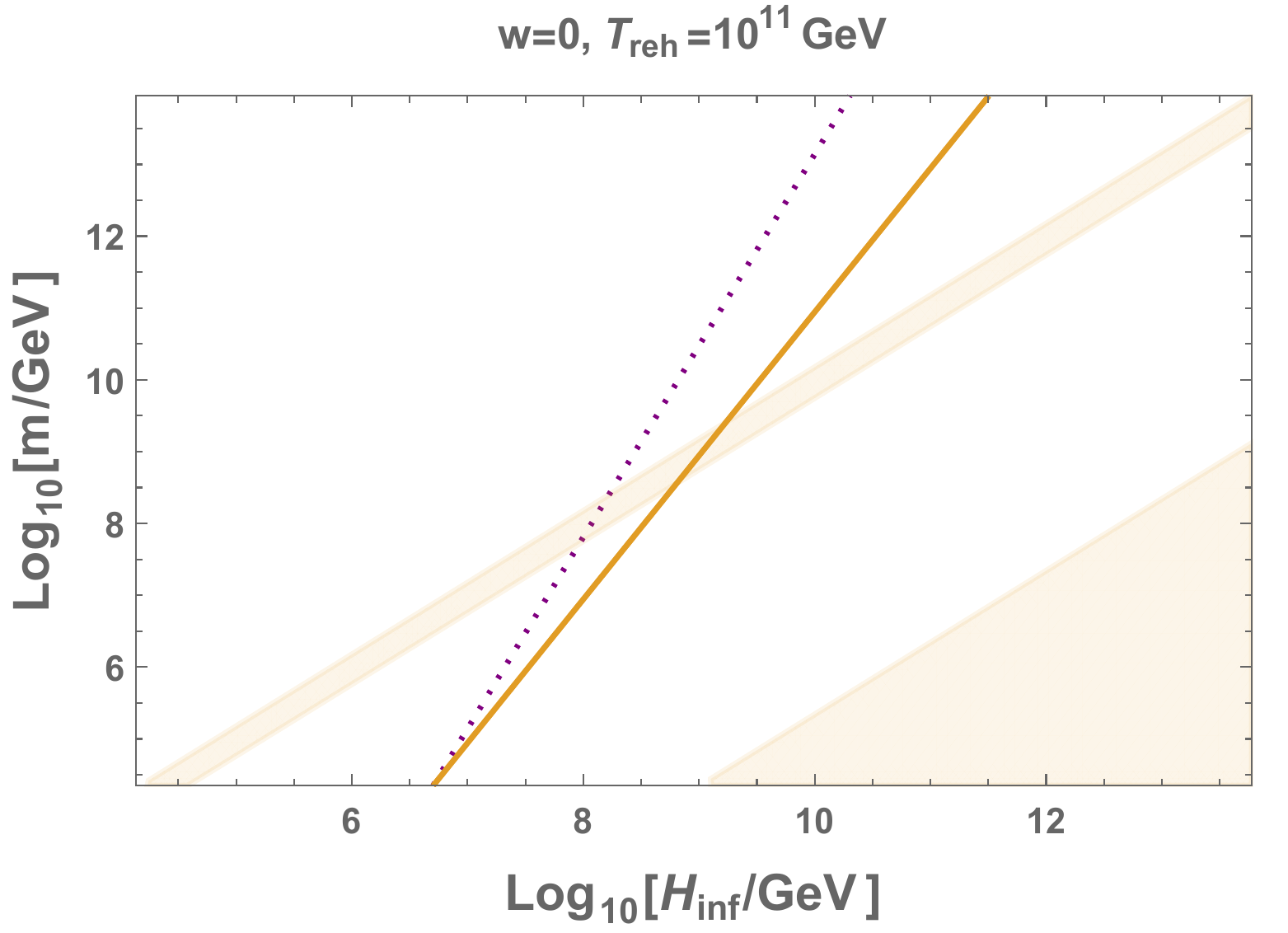}\vspace{0.5 cm}
\includegraphics[width=0.48\textwidth]{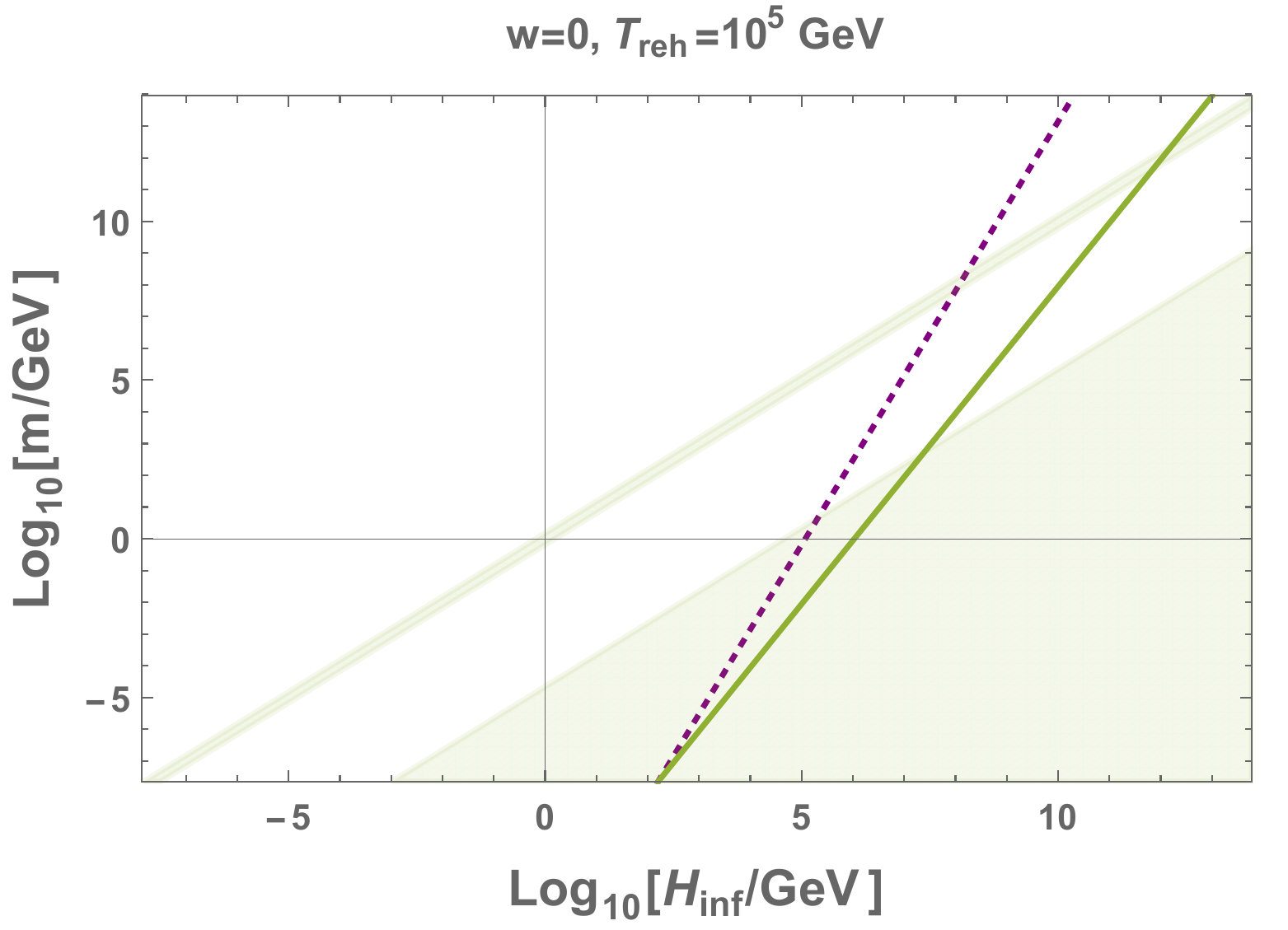}
\includegraphics[width=0.48\textwidth]{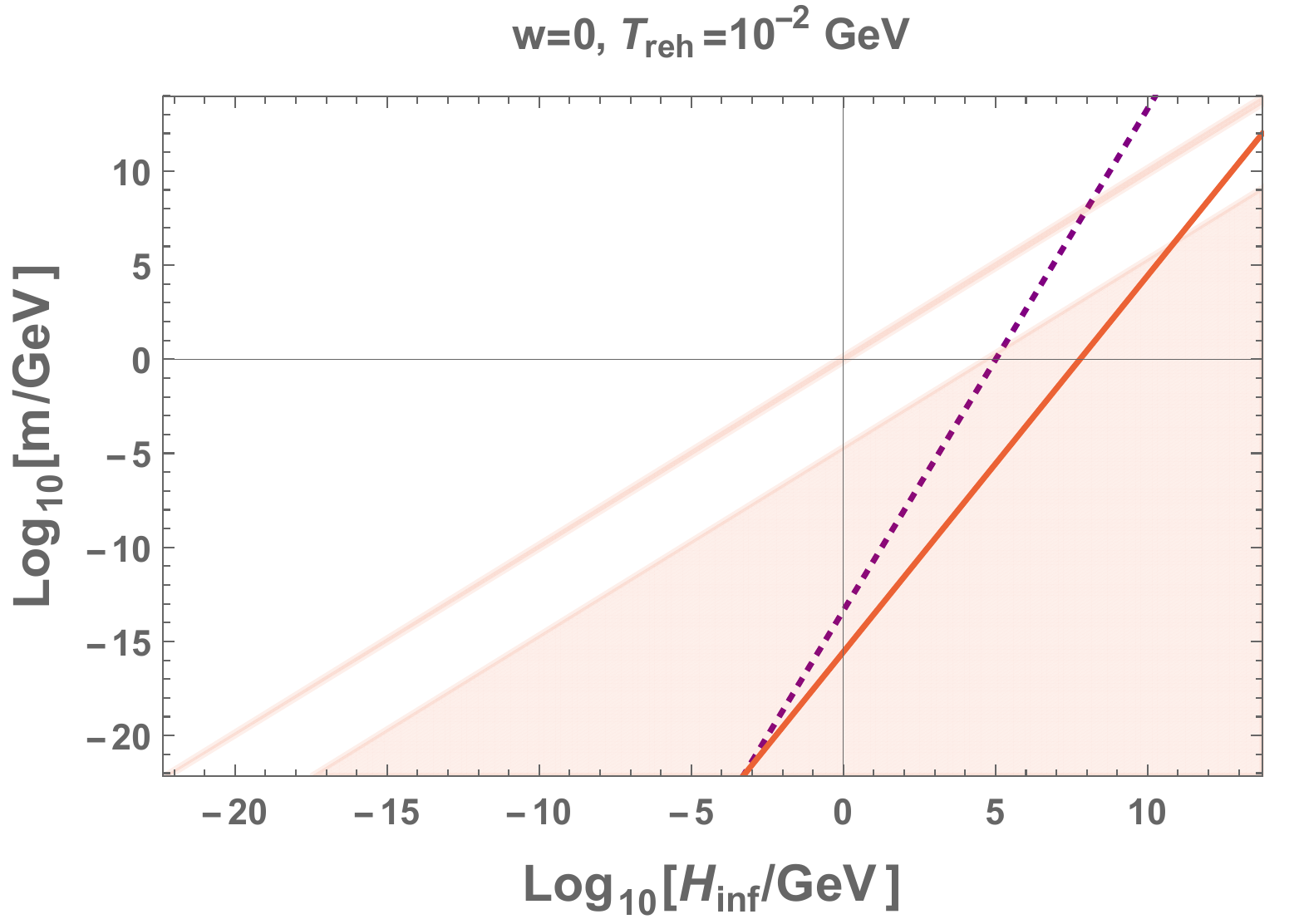}
\par\end{centering}
\caption{Parameter space of the model where the scalar constitutes all DM in the Universe (colored solid lines) and satisfies all constraints described in the beginning of Sec. \ref{results} (inside the shaded regions), assuming a quadratic potential and an equation of state parameter $w=0$. The DM abundance was computed for four different reheating temperatures as shown above each panel. The axis have been adjusted for each case to show only the part of the parameter space which is allowed by the constraints discussed in Sec. \ref{sec:constraints}. The dotted purple line corresponds to the DM abundance in the usual cosmological scenario with $w=1/3$, shown here for comparison. Note that the constraints shown here are for cases with $w=0$ and do not apply to the scenario with $w=1/3$ as such.
}
\label{w0 quad}
\end{figure}

\begin{figure}[H]
\begin{centering}
\includegraphics[width=0.48\textwidth]{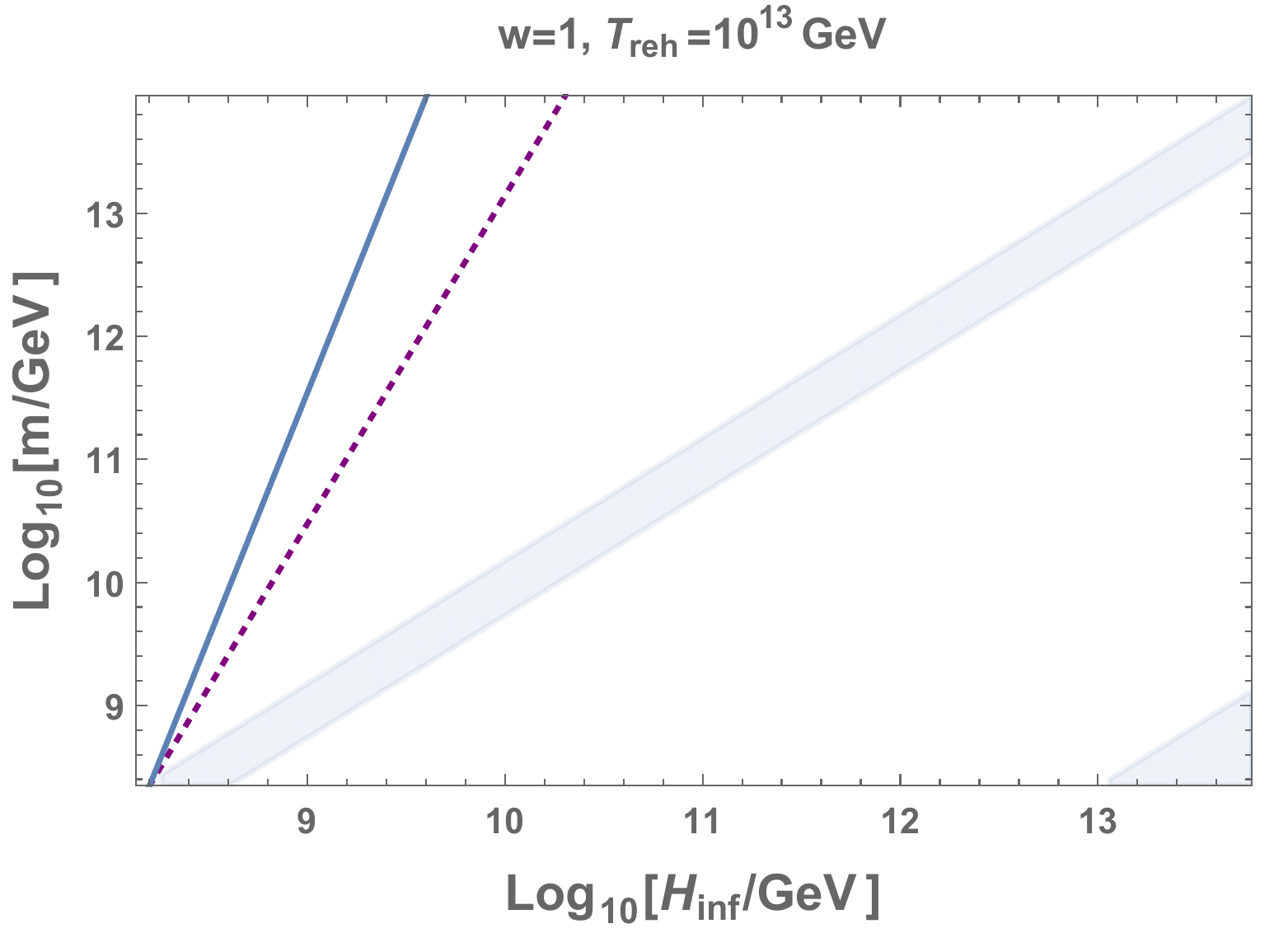}
\includegraphics[width=0.48\textwidth]{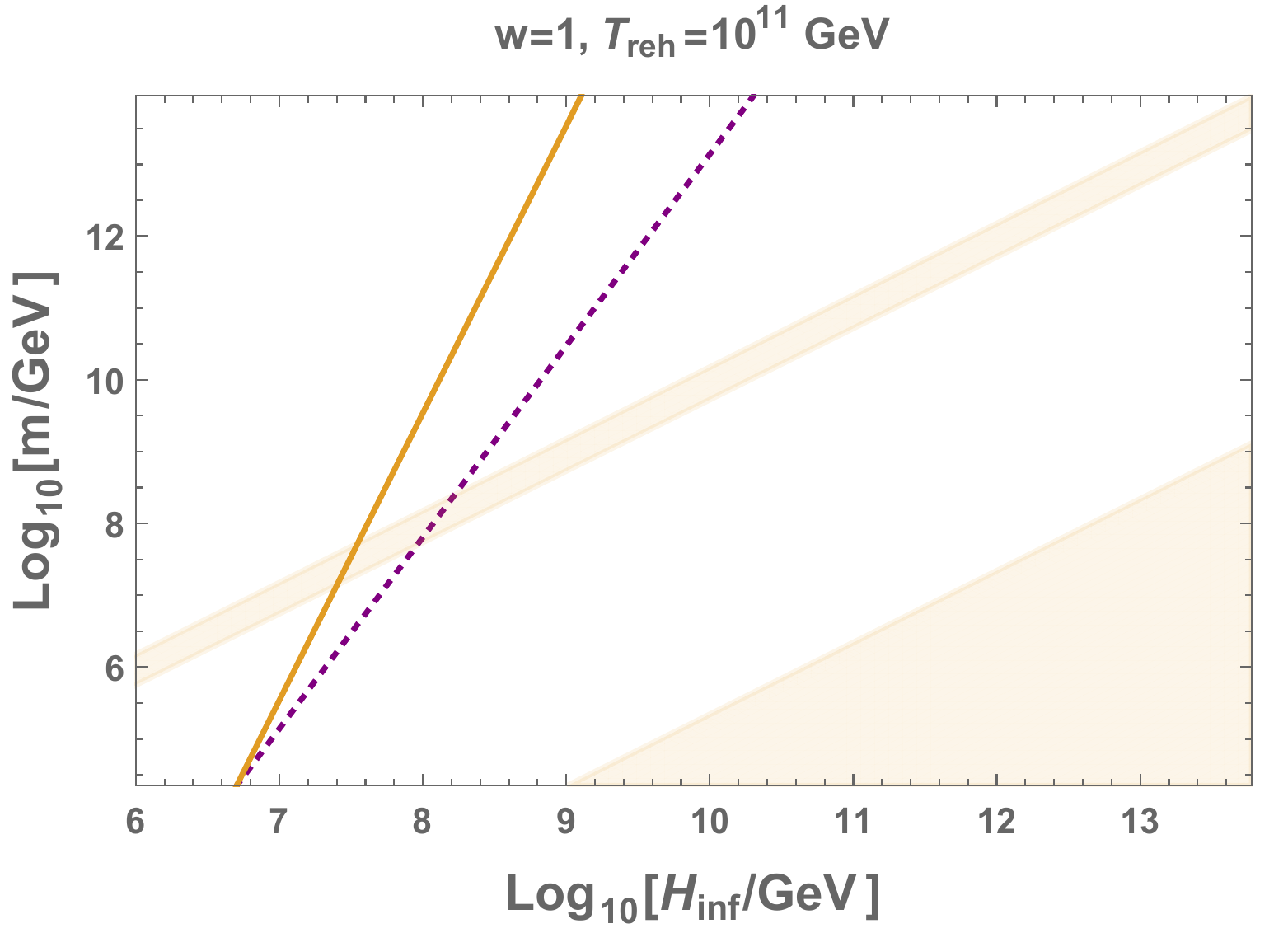}\vspace{0.5 cm}
\includegraphics[width=0.48\textwidth]{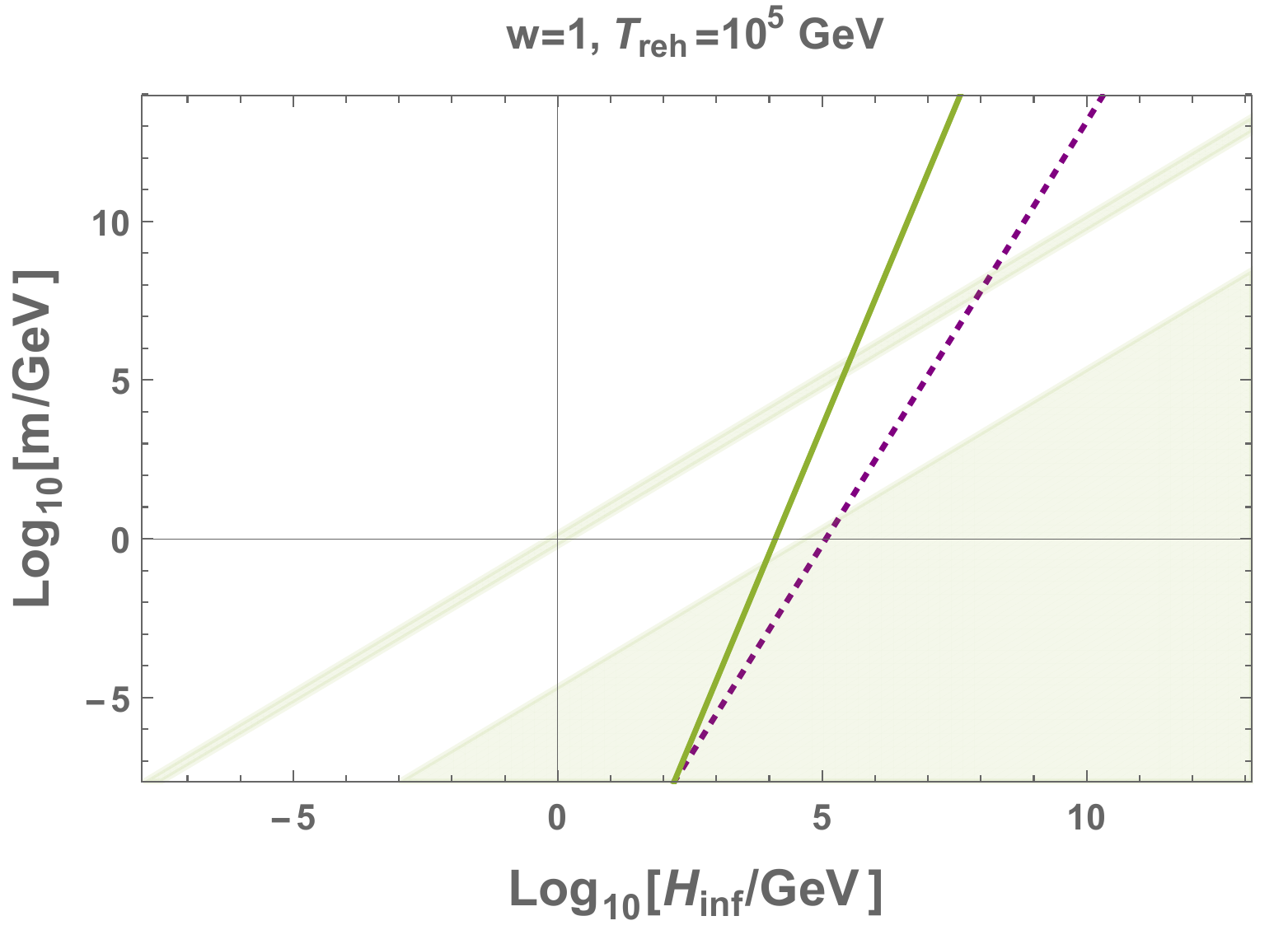}
\includegraphics[width=0.48\textwidth]{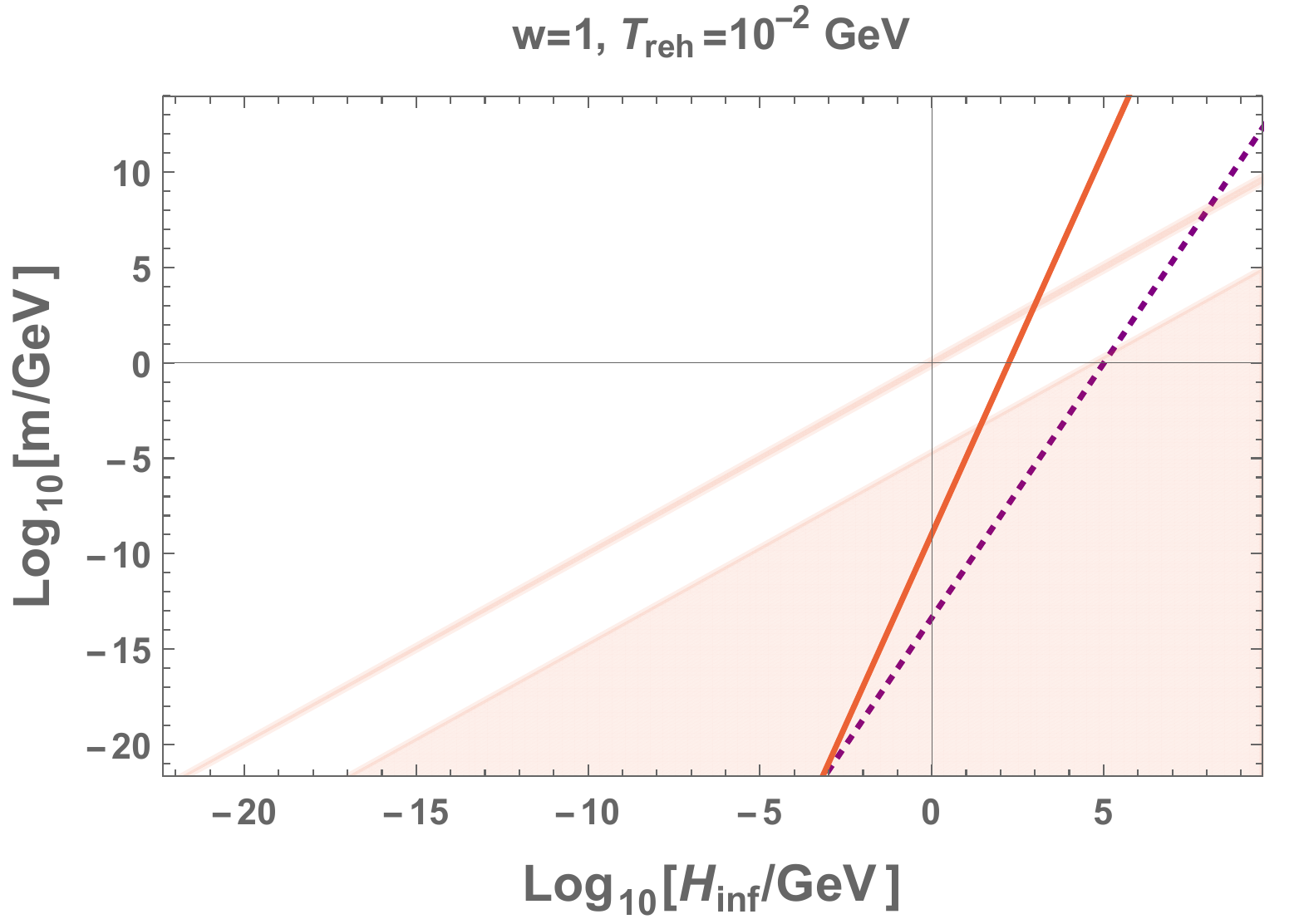}
\par\end{centering}
\caption{The same as Fig. \ref{w0 quad} but for $w=1$.
}
\label{w1 quad}
\end{figure}

First, Figs. \ref{w0 quad} and \ref{w1 quad} show the region of the model parameter space where the quadratic scenario ($m^2 > \lambda\chi_{\rm end}^2$) explains all DM (solid colored lines) and satisfies the constraints discussed in the beginning of this section (within the shaded regions) for the cases $w=0$ (Fig.~\ref{w0 quad}) and $w=1$ (Fig.~\ref{w1 quad}). We emphasize that as the shaded regions represent the part of the parameter space where the constraints are {\it satisfied}, the regions where the $\chi$ field can successfully explain all DM are those where the solid colored lines overlap with the shaded regions. 

As a few benchmark scenarios, we have considered four reheating temperatures,\linebreak $T_{\mathrm{reh}}=10^{13},\,10^{11},\,10^{5},\,10^{-2}$ GeV (shown in blue, yellow, green, and red, respectively), and assumed $g_{*}(T_{\mathrm{reh}})=100$ for all reheating temperatures except for $T_{\mathrm{reh}}=10^{-2}$ GeV, for which we used the more correct value $g_{*}(T_{\mathrm{reh}})=10$. The axis in each figure have been adjusted for each case to show only the part of the parameter space which is allowed by the constraints discussed in Sec. \ref{sec:constraints}. The dotted purple line corresponds to the DM abundance in the usual cosmological scenario with $w=1/3$ \cite{Tenkanen:2019aij}, which is shown here for comparison. Note, however, that this line shows only the DM abundance and the constraints shown in the plot do not apply to the scenario with $w=1/3$ as such but should be computed separately. Also, note that in all cases the DM isocurvature constraints have been computed assuming that the field constitutes all dark matter, and hence are applicable only along the solid lines.

We see that for fixed DM mass and reheating temperature, in the $w=0$ ($w=1$) case a higher (lower) value of $H_{\rm inf}$ than in the usual radiation-dominated scenario with $w=1/3$ is required to obtain the observed DM abundance today. This is naturally understood by the fact that the initial energy density stored in the spectator field is $\rho_\chi \propto H_{\rm inf}^4$, see Eq. \eqref{variance}, and the more the scalar field energy density becomes diluted compared to the background energy density after inflation, the higher the initial energy density of the scalar field (the value of $H_{\rm inf}$) has to be to obtain the correct DM abundance today. Note, however, that not all values of $w\,, T_{\rm reh}$ give the correct relic abundance for each set of $m\,,H_{\rm inf}$ shown in Figs.  \ref{w0 quad} and \ref{w1 quad}. For example, for $T_{\rm reh}=10^{13}$ GeV there is essentially no available parameter space where the $\chi$ field has the correct DM abundance today and simultaneously satisfies both the observational constraints and also the consistency conditions discussed in the previous subsection, neither for $w=0$ nor $w=1$. Here this scenario is shown as a limiting example to highlight the volume of the allowed ($w\,, T_{\rm reh}\,,m\,,H_{\rm inf}$) parameter space.

\begin{figure}[H]
\begin{centering}
\includegraphics[width=0.48\textwidth]{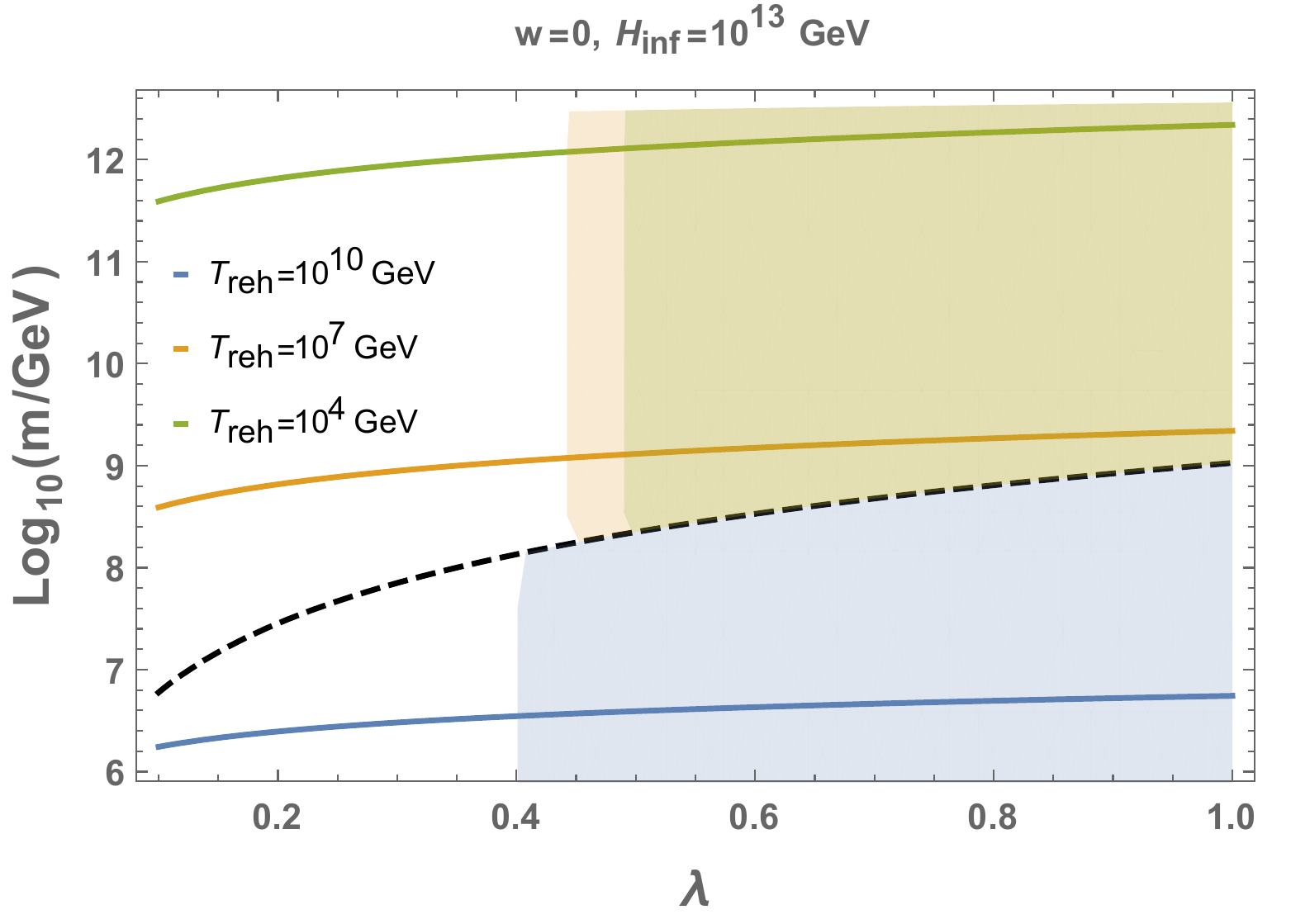}
\includegraphics[width=0.48\textwidth]{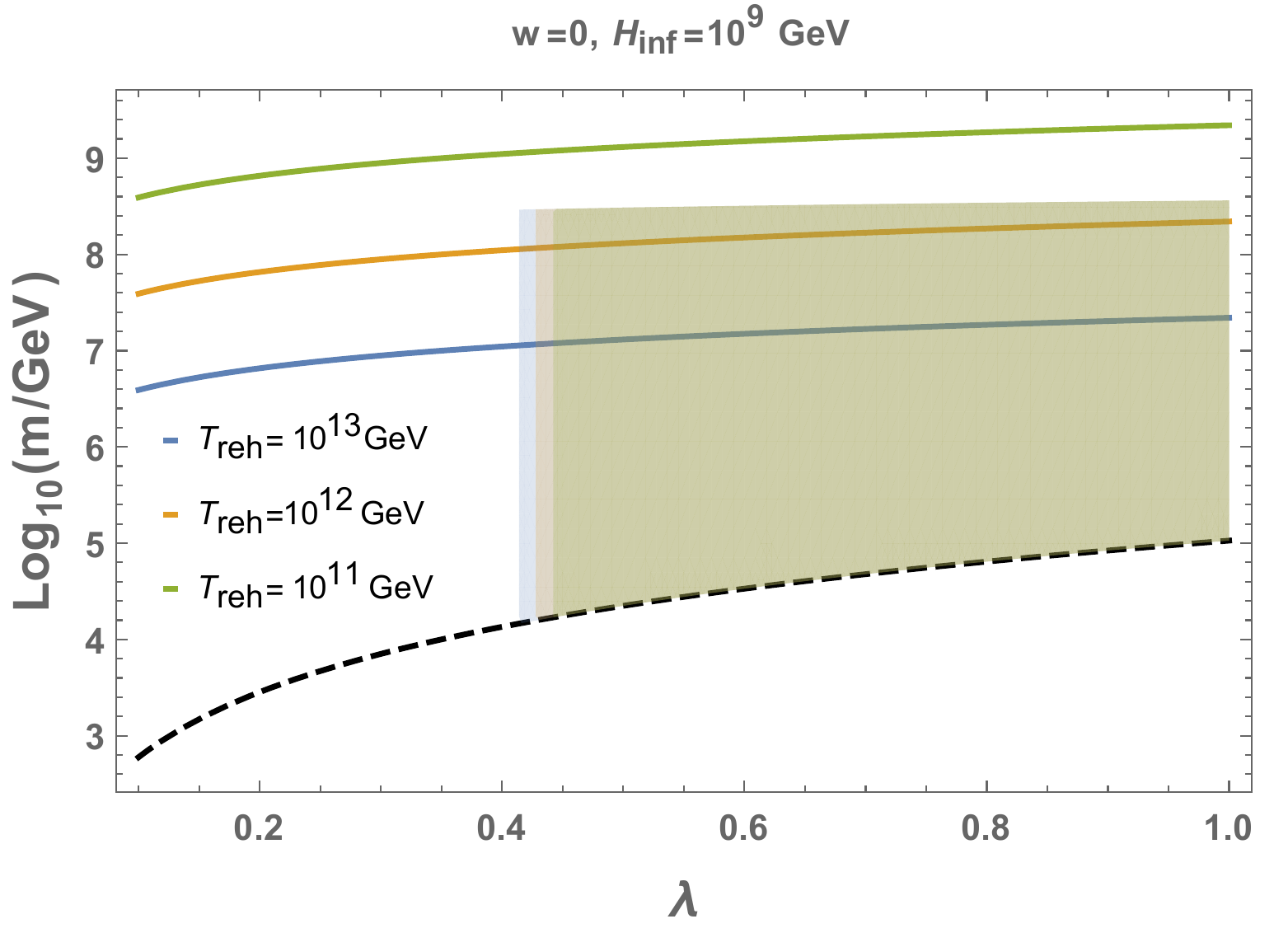}
\par\end{centering}
\caption{The parameter space of the model where the scalar constitutes all DM (colored solid lines) and simultaneously satisfies all constraints (inside the shaded regions), assuming a quartic potential and an equation of state parameter $w=0$. The DM abundance was computed for different reheating temperatures, as indicated in the plots. As two benchmark scenarios, we present those with $H_{\rm inf}=10^{13}$ GeV and $H_{\rm inf}=10^{9}$ GeV. Above the black dashed line the condensate remains coherent, while for masses below the dashed line the condensate evaporates. The axis have been adjusted for each case to show only the part of the parameter space which is allowed by the constraints discussed in Sec. \ref{sec:constraints}.
}
\label{w0 quar}
\end{figure}

\begin{figure}[H]
\begin{centering}
\includegraphics[width=0.48\textwidth]{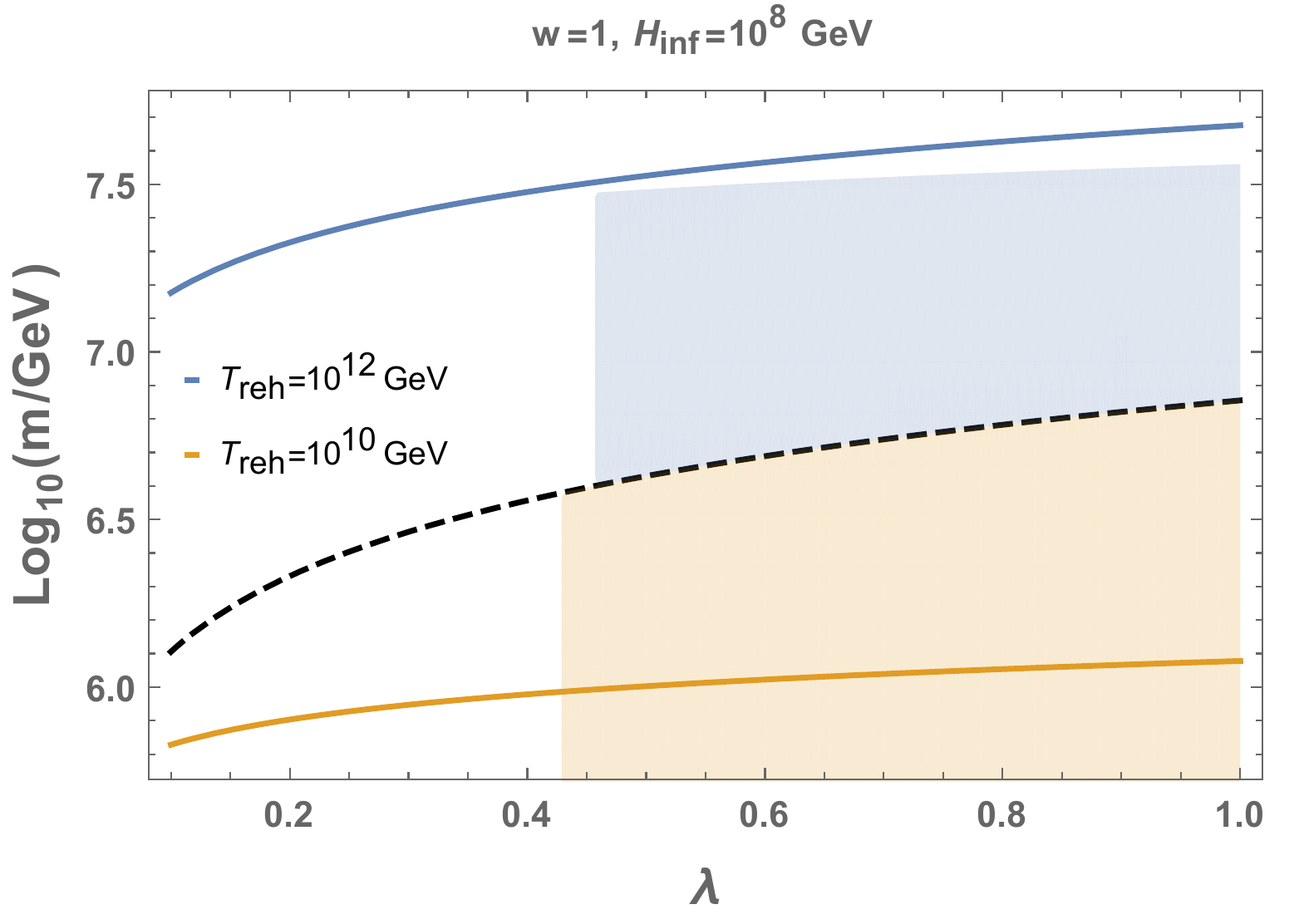}
\includegraphics[width=0.48\textwidth]{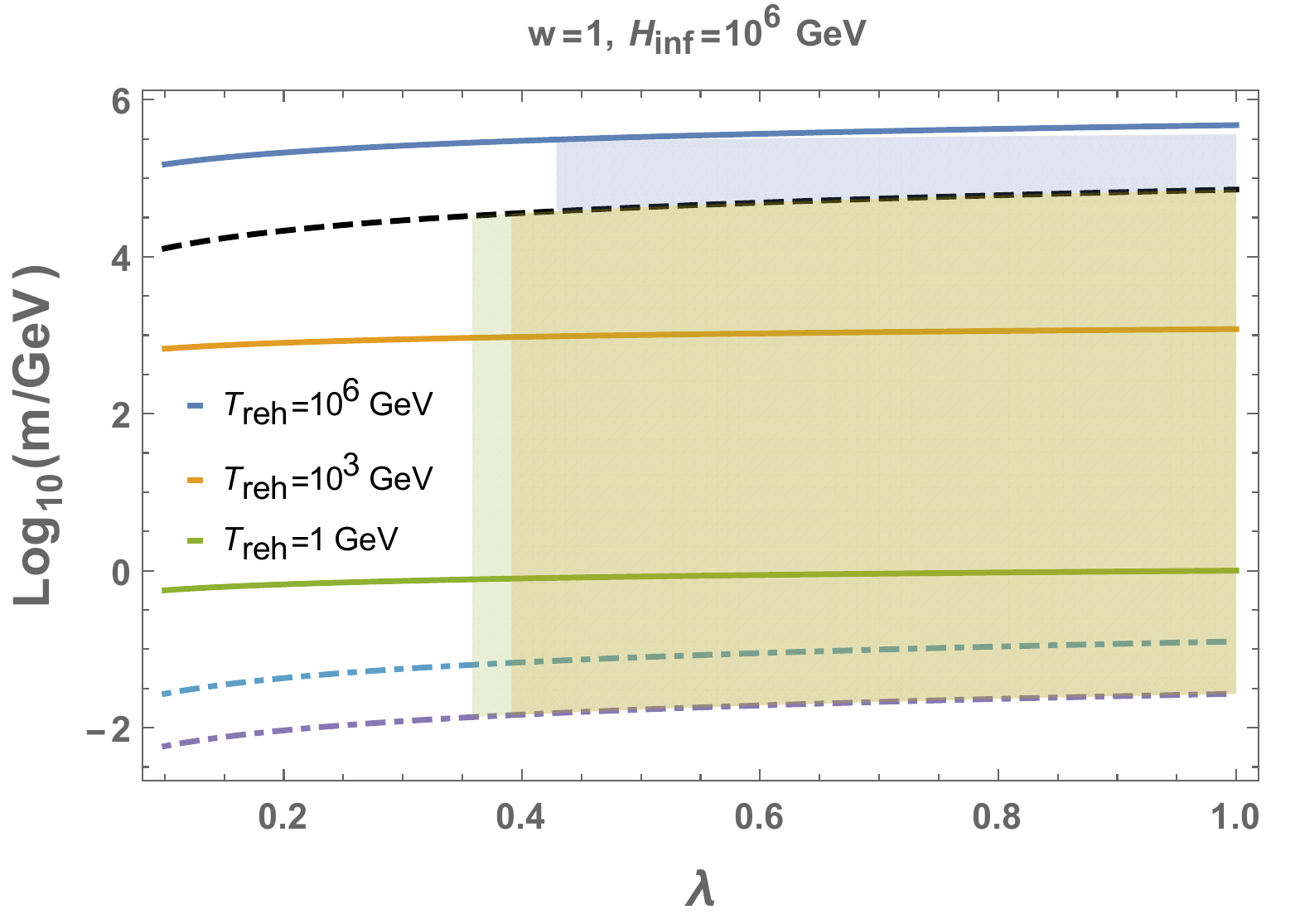}
\par\end{centering}
\caption{The same as Fig. \ref{w0 quar} but for $w=1$, $H_{\rm inf}=10^{8}$ GeV, and $H_{\rm inf}=10^{6}$ GeV. The dash-dotted lines on the right panel correspond to constraints on the DM self-interactions that can be inferred from galaxy clusters, Eq. \eqref{eq:sigmaDMbound}. The purple dash-dotted curve corresponds to $\sigma/m= 1\,{\rm cm}^2{\rm g}^ {-1}$, while the light blue one assumes $\sigma/m=10^{-2}\,{\rm cm}^2{\rm g}^ {-1}$, shown here as a target for future observations.
}
\label{w1 quar}
\end{figure}

In Figs. \ref{w0 quar} and \ref{w1 quar}, we present the allowed parameter space for the quartic scenario ($m^2 <\lambda\chi_{\mathrm{end}}^2$) in the $(\lambda,m)$ plane by fixing $H_{\mathrm{inf}}$ and varying $T_{\mathrm{reh}}$ for both $w=0$ and $w=1$. For the same reason as in the quadratic case, for fixed values of other parameters, in the $w=0$ ($w=1$) case a higher (lower) value of $H_{\rm inf}$ than in the usual radiation-dominated scenario with $w=1/3$ is required to obtain the observed DM abundance today. However, as the dimension of the parameter space is now greater than in the quadratic case, there are more combinations of parameters that allow the spectator field to successfully constitute all DM, as visualized in Figs. \ref{w0 quar} and \ref{w1 quar}. Note again that not all values of parameters shown there allow the field to constitute all DM; for example, in the right panel of Fig. \ref{w0 quar}, the case with $T_{\rm reh}=10^{11}$ GeV is ruled out. Another example is given in the right panel of Fig. \ref{w1 quar}, where the case with $T_{\rm reh}=10^{6}$ GeV is in tension with the constraints on our scenario. Note that in that figure, we also show the constraints on DM self-interactions. The purple dash-dotted curve at the bottom of the figure is a hard limit, depicting $\sigma/m= 1\,{\rm cm}^2{\rm g}^ {-1}$, whereas the blue curve assumes $\sigma/m=10^{-2}\,{\rm cm}^2{\rm g}^ {-1}$ and is shown here as a target for future observations (see Sec. \ref{testability}).

\subsection{Testability of the scenario}
\label{testability}

Finally, we discuss how the scenarios we have studied in this paper could be further constrained -- or supported -- by future observations. 

The fact that the DM perturbations in our scenario are genuinely of isocurvature type provides probably the best avenue for testing the scenario. First, the DM isocurvature perturbations generically enhance the perturbations in the SM matter density and can lead to a sizeable enhancement in the CMB temperature and/or matter power spectrum compared to the adiabatic case \cite{Graham:2015rva,Alonso-Alvarez:2018tus,Tenkanen:2019aij}. This is due to the fact that in the presence of isocurvature, the total curvature perturbation at superhorizon scales is given by (see e.g. Ref. \cite{Wands:2000dp})
\begin{equation}
\label{zeta_tot}
\zeta = \zeta_{\rm r} + \frac{z_{\rm eq}/z}{4+3z_{\rm eq}/z}S_{\rm r\chi}\,,
\end{equation}
where $\zeta_{\rm r}$ is the contribution of the SM radiation to the curvature perturbation and $z$ ($z_{\rm eq}$) is the redshift (to the matter-radiation equality), so that at late times the prefactor of the second term is $\sim 1/3$. In our scenario deviations from the adiabatic case can be large especially at smaller physical distance scales, which is due to the fact that in our scenario the DM isocurvature spectrum is always blue-tilted (see Fig. \ref{fig:P_S}). At subhorizon scales the simple picture of Eq. \eqref{zeta_tot} does not hold but one can nevertheless show that also at those scales the effect of DM isocurvature is to increase the curvature perturbation at the linear level (see e.g. Ref. \cite{Feix:2020txt}). Furthermore, because in e.g. the usual axion DM models the corresponding power spectrum is typically nearly scale-invariant \cite{Beltran:2006sq,Visinelli:2009zm}, the spectator DM model may be distinguishable from this type of models if observations of the matter power spectrum can be extended to large enough $k$ where the data may show evidence for (scale-dependent) DM isocurvature; a potentially promising new avenue in this respect is the forthcoming Euclid satellite mission \cite{Laureijs:2011gra,Amendola:2016saw}. It is worth noting here that some recent analysis of the CMB temperature fluctuations have actually shown hints of a blue-tilted DM isocurvature contribution \cite{Chung:2017uzc,Akrami:2018odb,Feix:2020txt}, although some of these results are still preliminary and the effects of isocurvature are degenerate with the effect of other cosmological parameters.

Second, the isocurvature nature of DM in our scenario provides also an additional way to both distinguish our scenario from other models and also further test the properties of DM. That is, due to the stochastic behavior of the $\chi$ field during inflation, in all cases studied in this paper the DM isocurvature is non-Gaussian, which in practice shows not only in the form of the three-point correlation function of the DM isocurvature perturbation, $\langle S_{\rm r\chi}(\bar{x}_1)S_{\rm r\chi}(\bar{x}_2)S_{\rm r\chi}(\bar{x}_3)\rangle$ (where $S_{\rm r\chi}$ is given by Eq. \eqref{Srchi}), but also in the three-point correlator of the total curvature perturbation, $\langle \zeta(\bar{x}_1)\zeta(\bar{x}_2)\zeta(\bar{x}_3)\rangle$, which is partly determined by the former (see Eq. \eqref{zeta_tot}). Because the perturbations are uncorrelated, we have
\begin{equation}
\label{NG}
\langle \zeta(\bar{x}_1)\zeta(\bar{x}_2)\zeta(\bar{x}_3)\rangle = \langle \zeta_{\rm r}(\bar{x}_1)\zeta_{\rm r}(\bar{x}_2)\zeta_{\rm r}(\bar{x}_3)\rangle + \left(\frac{z_{\rm eq}/z}{4+3z_{\rm eq}/z}\right)^3\langle S_{\rm r\chi}(\bar{x}_1)S_{\rm r\chi}(\bar{x}_2)S_{\rm r\chi}(\bar{x}_3)\rangle\,,
\end{equation}
where the prefactor of the second term is $\sim 1/27$ at the time of last scattering and the three-point correlator for $\zeta_{\rm r}$ depends on the inflationary model \cite{Maldacena:2002vr} (and possibly also the fluctuations the SM Higgs acquired during inflation \cite{Tenkanen:2019cik}). Therefore, even if the non-Gaussianity generated during inflation was negligible, DM isocurvature perturbations can generate sizeable non-Gaussianity at late times. In the spectator DM scenario the effect is again different from the usual axion DM scenario and also many other models where non-Gaussianity can be generated, such as curvaton models (see e.g. Refs. \cite{Kawasaki:2008sn,Langlois:2008vk,Langlois:2011hn,Langlois:2012tm,Hikage:2012be,Kitajima:2017fiy}), as the non-Gaussianity in the present scenario is of non-local type\footnote{Local non-Gaussianity is defined as $\zeta(\bar{x}) = \zeta_{\rm G}(\bar{x})+\frac{3}{5}f_{\rm NL}\left(\zeta_{\rm G}^2(\bar{x})-\langle\zeta_{\rm G}^2(\bar{x})\rangle\right)$, where $f_{\rm NL}$ is the first order non-Gaussianity parameter and $\zeta_{\rm G}$ denotes the Gaussian part of the curvature perturbation (see e.g. Ref. \cite{Byrnes:2014pja}). In our scenario the curvature perturbation does not take the local form and is hence ``non-local".}. While computing the above three-point correlation functions is beyond the scope of this paper, doing so would certainly be worthwhile if primordial DM isocurvature or non-Gaussianity were discovered, as they may provide a powerful way to test the spectator DM scenario and distinguish it from other models, such as those where axions constitute the DM\footnote{For an early work that computed the three-point correlator of a stochastic scalar field in the special case of equilateral triangles, see Ref. \cite{Peebles:1999fz}. It would be interesting to generalize this calculation to other shapes as well.}.

Finally, we comment on the prospects for detecting DM self-interactions. While in the quadratic scenario studied in this paper the DM self-interactions are by definition negligible, in the opposite case where the quartic term dominated the scalar field potential already during inflation, the DM self-interactions can be sizeable. Currently observations of dynamics of celestial bodies at the galactic and galaxy cluster scales place an upper bound on the DM self-interaction cross-section over DM mass which is of the order $\sigma/m\leq 1$ cm$^2$/g, however, in the future this limit may be possible to become tightened to $\sigma/m\lesssim \mathcal{O}(0.01)$ cm$^2$/g \cite{Tulin:2017ara}. As our results show (see Fig. \ref{w1 quar}), this will further constrain the model parameter space in the quartic case. In case of a positive detection, the quartic case can accommodate these interactions for suitable $m$ and $\lambda$ (as well as $T_{\rm reh}\, H_{\rm inf}, w$), while the quadratic case would obviously be ruled out. It should be noted that this is in contrast to the case with standard cosmology, where the spectator DM model can never account for DM self-interactions of observable size \cite{Markkanen:2018gcw}. As we have shown in this paper, however, with a modified cosmological history this is not a problem.

%%%%%%%%%%%%%%%%%%%%%%%%%%%%%%%%%%%%%%%%%%%%%%%%%%%%%%%%%%%%%%%%%%%%%%%%%%%%%%%%%%%%%%%%%%

\section{Conclusions}
\label{conclusions}

In this paper, we have studied the spectator dark matter scenario, where the observed DM abundance is produced by amplification of quantum fluctuations of an energetically subdominant scalar field during inflation. We showed that the scenario is robust to changes in the expansion history of the early Universe in a sense that also in this case the scenario works for a wide range of DM masses and coupling values, although even relatively modest changes to the standard cosmological history can impose notable quantitative differences to the usual scenario. This is because in the presence of a non-standard expansion phase the DM energy density evolves differently as a function of time, and also the DM isocurvature perturbation spectrum turns out to be different from the result in the radiation-dominated case. We quantified these differences in both free and self-interacting DM cases and presented the refined model parameter space which allows the scalar field to constitute all DM while simultaneously satisfying all observational constraints. 

While we have discussed only few example cases (early matter-domination and kination-domination encountered in e.g. quintessence models), further modifications to the early cosmological history can also be imagined. Likewise, it would be interesting to see how a non-standard phase of expansion in the early Universe can change the allowed model parameter space in scenarios where the DM field couples non-minimally to gravity, to the inflaton field, and/or to another spectator field, for instance the SM Higgs.

Finally, we discussed the prospects for testing the scenario with future observations. In particular, if primordial DM isocurvature or non-Gaussianity is ever discovered, this may provide a powerful way to test the spectator DM scenario and distinguish it from other models, such as those where axions constitute the DM. Indeed, it is worth emphasizing that despite the fact that in these models the DM field interacts with ordinary matter only via gravity, these scenarios are testable with both current and future observations of the CMB and the large scale structure of the Universe, as well as the dynamics of celestial bodies at galactic and galaxy cluster scales, as discussed in this paper. If observations ever show any deviation from the adiabatic, non-interacting cold DM paradigm, it would be interesting to see what that tells about DM candidates which are only gravitationally interacting. After all, for all we know about dark matter, this minimal scenario seems to be the one preferred by observations.

%%%%%%%%%%%%%%%%%%%%%%%%%%%%%%%%%%%%%%%%%%%%%%%%%%%%%%%%%%%%%%%%%%%%%%%%%%%%%%%%%%%%%%%%%%

\acknowledgments
We thank D. Bettoni, M. Kamionkowski, A. Rajantie, and J. Rubio for correspondence and discussions. C.C. is supported by the Arthur B. McDonald Canadian Astroparticle Physics Reasearch Institute and T.T. by the Simons foundation. T.T. thanks the hospitality of Carleton University, where this work was initiated.

%%%%%%%%%%%%%%%%%%%%%%%%%%%%%%%%%%%%%%%%%%%%%%%%%%%%%%%%%%%%%%%%%%%%%%%%%%%%%%%%%%%%%%%%%%

\appendix

\section{The Klein-Gordon equation with non-standard expansion at early times}
\label{appendix}

Here we present the equation of motion for a homogeneous, conformal scalar field ($V(\chi)=\lambda/4 \chi^4$) in the general case ($-1/3 < w \leq 1$).

The Klein-Gordon equation for the scalar field is
\begin{equation}
\label{app_KG}
\ddot{\chi} + 3H(t)\dot{\chi} + \lambda\chi^3 = 0\,,
\end{equation}
where $H(t)$ is determined by the background energy density and its time-dependence, which in turn is determined by the equation of state parameter $w$. Upon changing to conformal time ${\rm d}\eta = {\rm d}t/a$ and defining $z\equiv a\sqrt{\lambda}\chi$, Eq. (\ref{app_KG}) becomes
\begin{equation}
\label{rescaled_KG}
z'' + \left[\frac12\left(1+3w\right)-1\right]\mathcal{H}^2 z + z^3 = 0\,,
\end{equation}
where $\mathcal{H} \equiv a'/a$ is the conformal Hubble parameter, the prime denotes derivative with respect to $\eta$, and we used the well-known result $\mathcal{H}' = -\frac12(1+3w)\mathcal{H}^2$.

\begin{figure}[httb]
\begin{centering}
\includegraphics[scale=.6]{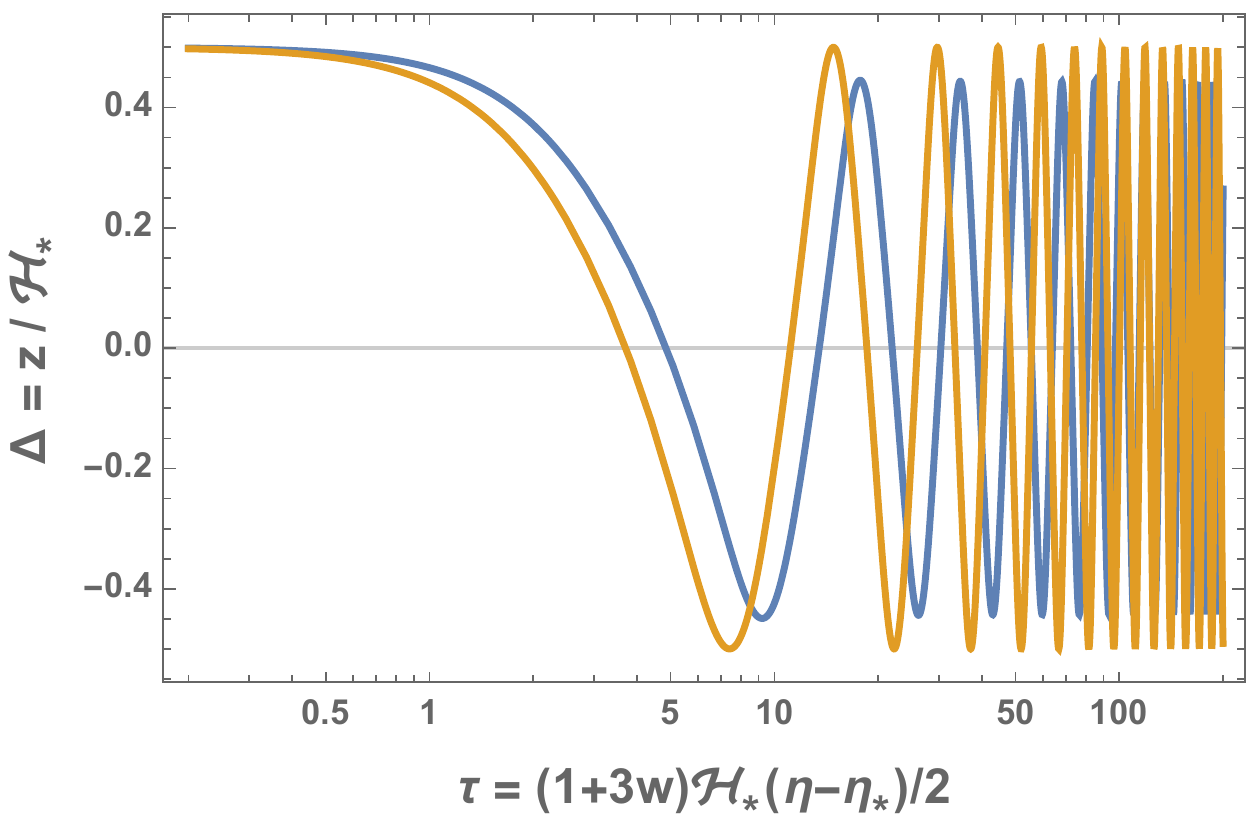}
\includegraphics[scale=.6]{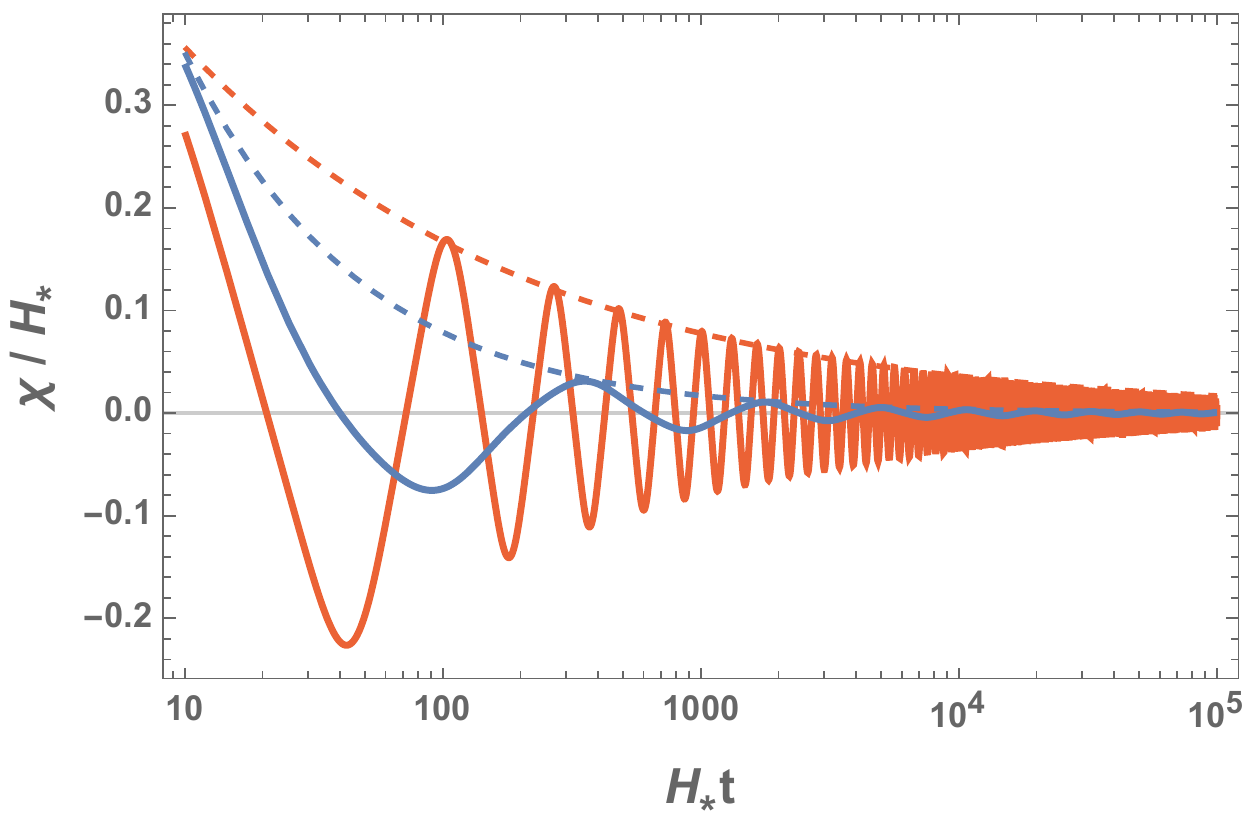}
\par\end{centering}
\caption{{\it Left panel}: Solutions to Eq. \eqref{DeltaEq} in rescaled conformal time $\tau$ in the $w=1/3$ (orange) and $w=0$ (blue) cases. {\it Right panel}: Solutions to the scalar field equation of motion \eqref{app_KG} in cosmic time $t$ in the $w=0$ (blue) and $w=1$ (red) cases with radiation-like $\chi \propto 1/a$ scaling superimposed by the dashed curves, which the oscillation envelopes very precisely follow in both cases. In this figure $\lambda = 0.1$.
 \label{fig:KG}}
\end{figure}

In conformal time, the Friedmann equation reads 
\begin{equation}
3\mathcal{H}^2 M_{\rm P}^2 = \rho_{\rm bg} a^2\,,
\end{equation}
where "bg" stands for background. Here $\rho_{\rm bg} \propto a^{-3(1+w)}$ as usual. Thus, we obtain
\begin{equation}
\mathcal{H} = \frac{\mathcal{H}_*}{\frac{1+3w}{2} H_*(\eta-\eta_*) + 1}\,,
\end{equation}
where $\mathcal{H}_* \equiv \mathcal{H}(\eta=\eta_*)$, where the subscripts denote the initial time. Thus, the general form for the scalar field equation of motion (\ref{rescaled_KG}) is
\begin{equation}
\label{app_rescaled_eom}
z'' + F(\eta, w)z + z^3 = 0\,,
\end{equation}
where
\begin{equation}
\label{Fterm}
F(\eta, w) \equiv \frac{\left[\frac12(1+3w)-1\right]\mathcal{H}_*^2}{\left[\frac{1+3w}{2}\mathcal{H}_*(\eta-\eta_*)+1\right]^2}\,.
\end{equation}
Clearly, when $w=1/3$, Eq. (\ref{app_rescaled_eom}) reduces to
\begin{equation}
z'' + z^3 = 0\,,
\end{equation}
whose solution is a well-known oscillating function: the elliptic (Jacobi) cosine, whose exact form can be found analytically (see e.g. Refs. \cite{Ichikawa:2008ne,Kainulainen:2016vzv}) and which, besides the oscillations, has no further time-dependence in terms of $\eta$ (see Fig. \ref{fig:KG}). Because $\chi \propto z/a$, this means that when the background energy density is radiation-dominated, the oscillation amplitude decays simply as $\chi \propto 1/a$, i.e. as that of (dark) radiation.

The above conclusion applies to the other cases as well. Upon rescaling $\tau \equiv (1+3w)/2\mathcal{H}_*(\eta-\eta_*)$ and $\Delta \equiv z/\mathcal{H}_*$, we obtain
\begin{equation}
\label{DeltaEq}
\Delta'' + \frac{p}{(1+\tau)^2}\Delta + q\Delta^3 = 0\,,
\end{equation}
where the primes now denote derivative with respect to $\tau$ and $p,q$ are numbers which depend on $w$ and the normalization of $\mathcal{H}_*,\eta_*$ and whose exact expressions are irrelevant here. The form of Eq. \eqref{DeltaEq} is particularly useful for numerical analysis, as all quantities are dimensionless and expressed in Hubble units (units of $\mathcal{H}_*$). The numerical analysis shows that the term $F$-term (\ref{Fterm}) dies off quickly and thus in all cases we retain the usual $\chi \propto 1/a$ scaling, which validates the treatment of the scalar field energy density in Sec. \ref{sec:quartic}.

%%%%%%%%%%%%%%%%%%%%%%%%%%%%%%%%%%%%%%%%%%%%%%%%%%%%%%%%%%%%%%%%%%%%%%%%%%%%%%%%%%%%%%%%%%

%\clearpage 

\bibliography{ScalarDM}

%%%%%%%%%%%%%%%%%%%%%%%%%%%%%%%%%%%%%%%%%%%%%%%%%%%%%%%%%%%%%%%%%%%%%%%%%%%%%%%%%%%%%%%%%%

\end{document}